\documentclass[11pt]{article}
\usepackage[latin9]{inputenc}
\usepackage{amsmath}
\usepackage{amssymb}
\usepackage{graphicx}
\usepackage{esint}
\usepackage[unicode=true,
 bookmarks=false,
 breaklinks=false,pdfborder={0 0 1},backref=section,colorlinks=false]
 {hyperref}

\makeatletter

\providecommand{\tabularnewline}{\\}

\@ifundefined{date}{}{\date{}}

\usepackage{esint}
\usepackage{latexsym}
\usepackage{graphicx}\usepackage{bm}\usepackage{longtable}

\usepackage{xcolor}

\@addtoreset{equation}{section}

\makeatother

\begin{document}
\title{Holographic Schwinger effect in the confining background with D-instanton}
\maketitle
\begin{center}
\footnote{Email: siwenli@dlmu.edu.cn}Si-wen Li\emph{$^{\dagger}$}
\par\end{center}

\begin{center}
\emph{$^{\dagger}$Department of Physics, School of Science,}\\
\emph{Dalian Maritime University, }\\
\emph{Dalian 116026, China}\\
\par\end{center}

\vspace{8mm}

\begin{abstract}
Using the gauge-gravity duality, we study the holographic Schwinger
effect by performing the potential analysis on the confining D3- and
D4-brane background with D-instantons then evaluate the pair production/decay
rate by taking account into a fundamental string and a single flavor
brane respectively. The two confining backgrounds with D-instantons
are obtained from the black D(-1)-D3 and D0-D4 solution with a double
Wick rotation. The total potential and pair production/decay rate
in the Schwinger effect are calculated numerically by examining the
NG action of a fundamental string and the DBI action of a single flavor
brane all in the presence of an electric field. In both backgrounds
our numerical calculation agrees with the critical electric field
evaluated from the DBI action and shows the potential barrier is increased
by the presence of the D-instantons, thus the production/decay rate
is suppressed by the D-instantons. Our interpretation is that particles
in the dual field theory could acquire an effective mass through the
Chern-Simons interaction or the theta term due to the presence of
D-instantons so that the pair production/decay rate in Schwinger effect
is suppressed since it behaves as $e^{-m^{2}}$. Our conclusion is
in agreement with the previous results obtained in the deconfined
D(-1)-D3 background at zero temperature limit and from the approach
of the flavor brane in the D0-D4 background. In this sense, this work
may be also remarkable to study the phase transition in Maxwell-Chern-Simons
theory and observable effects by the theta angle in QCD.
\end{abstract}
\newpage{}

\section{Introduction}

People have achieved many advances in the researches about the phenomena
with strong electromagnetic field in the heavy-ion collision. The
Schwinger effect as one of the most famous phenomenon attracts great
interests since it is significantly related to the particle creation
rate. Specifically the charged particles in the collisions at high
speed can generate an extremely strong electromagnetic field so that
the virtual pairs of particles in the vacuum can become real particles
\cite{Schwinger,Affleck}. Thus the Schwinger effect may be very helpful
to study particle creation and thermalization in the heavy-ion collision.
On the other hand, the P or CP violation in QCD (Quantum Chromodynamics)
is also an important topic \cite{Phys.Rept theta}. Usually a theta
term can be added to the gauge theory to include the P or CP violation
in the action,

\begin{equation}
S=-\frac{1}{2g_{YM}^{2}}\mathrm{Tr}\int F\wedge^{*}F+i\frac{\theta}{8\pi^{2}}\mathrm{Tr}\int F\wedge F,\label{eq:1}
\end{equation}
where $g_{YM}$ refers to the Yang-Mills coupling constant. Although
the experimental value of the theta angle is very small $\left(\left|\theta\right|\leq10^{-10}\right)$,
the theta-dependence in Yang-Mills theory or QCD is very interesting
both in the theoretical and phenomenological researches, e.g. study
of the phase transition of confinement/deconfinement \cite{Massimo D'Elia - 1,Massimo D'Elia - 2},
the large $N$ behavior \cite{Witten theta}, the glueball spectrum
\cite{Luigi Del Debbio spectrum} all in the presence of a theta angle.
Particularly whether a theta vacua can be created in heavy-ion collision
is still an open question which therefore attracts many investigations
\cite{Kharzeev,Buckley,Kharzeev - 1,Kharzeev - 2} and moreover there
have been some observable effects proposed in order to confirm the
theta-dependence in the heavy-ion collision e.g. the chiral magnet
effect \cite{Kharzeev CME,Kharzeev CME - 2}.

Accordingly the motivation of this work is to explore the Schwinger
effect with the theta angle in the QCD-like or confining theory since
the Schwinger effect would be affected by the theta angle and it might
be remarkable to confirm the existence of the theta vacuum. However,
using QFT (quantum field theory) frame work to analyze the Schwinger
effect in a QCD-like or confining theory would be very challenging
since the original Schwinger\textquoteright s work shows that this
effect must be non-perturbative. Fortunately, the gauge-gravity duality
could provide an alternative way \cite{Maldacena ads/cft,Witten ads}
to analyze the Schwinger effect by evaluating the total potential
in holography \cite{Semenoff Schwinger effect}. In order to study
the QCD-like theory, the confining background geometry is also necessary.
Taking account into the frame work of string theory, the most famous
confining (soliton) background geometry is obtained by a double Wick
rotation on the black D3- or D4-brane solution \cite{Witten confinement,BBS book},
so the holographic total potential of Schwinger effect can be calculated
by performing the potential analysis \cite{Y. Sato} in these backgrounds
as in \cite{Confining D3 - 1,Confining D3 - 2}. To involve the theta
angle or the topological charge in the dual field theory, we additionally
need to introduce D-instantons into the D3- and D4-brane solution.
This can be achieved by considering the black D(-1)-D3 \cite{Liu hong D-1D3,Sangjin Sin,Li D instanton,Si-wen Li CS}
and D0-D4 brane solution \cite{Kenji Suzuki D0-D4,WuChao D0-D4} with
a double Wick rotation, following the discussion in \cite{Witten confinement,BBS book}.
Afterwards the D(-1) and D0-branes play the role of the D-instantons
and holographically correspond to the coupling constant of the Chern-Simons
term or the theta anglein the dual theory \cite{WuChao D0-D4,Sangjin Sin D0-D4,Francesco D0-D4 - 1,Francesco D0-D4 - 2,Li D0-D4}.

In this project, we will perform the potential analysis \cite{Y. Sato}
for the holographic Schwinger effect in the soliton D3- and D4-brane
background with D-instantons respectively, then evaluate the production/decay
rate by involving a fundamental string and a single flavor brane.
Our analysis shows the Schwinger effect occurs above a certain critical
electric field. The presence of the D-instantons increases the critical
electric field and the potential barrier of the Schwinger effect,
thus suppresses the production/decay rate both in the D(-1)-D3 and
D0-D4 background. The numerical calculation of the fundamental string
and the flavor brane consistently supports our analysis which might
imply the universal features of the Schwinger effect with instantons. 

The outline of this paper is as follows. In Section 2, we give a brief
construction for the confining D3- and D4-brane background with D-instantons.
In Section 3, we perform potential analysis for the holographic Schwinger
effect then calculate the total potential numerically. In Section
4, the pair production rate is evaluated with D-instantons. As a supplement
to this work, we analyze parallelly the Schwinger effect by taking
account into a single flavor in Section 5 and in Section 6, we give
our summary and discussion of this work. Basically our work is an
extension to \cite{Confining D3 - 1,Confining D3 - 2}, also a different
approach to check the results obtained in the deconfined D(-1)-D3
background at zero temperature limit \cite{deconfined D(-1)-D3} and
the flavor brane setup for holographic Schwinger effect in \cite{Wenhe}.

\section{The confining geometry with D-instanton}

\subsection{The confining D(-1)-D3 solution}

The D(-1)-D3 brane system was proposed in \cite{Liu hong D-1D3} which
is represented by a deformed D3-brane solution with a Ramond-Ramond
(RR) nontrivial scalar field. And the Ramond-Ramond (RR) scalar charge
is balanced by the dilaton in order to preserve 1/2 of supersymmetry.
This system is recognized as a marginal \textquotedblleft bound state\textquotedblright{}
of D3-branes with smeared D(-1)-branes , i.e. the D-instantons and
its low-energy dynamics is described by the type IIB supergravity
action. The non-extremal solution for the black D3-branes with D(-1)-branes
as D-instantons can be found in \cite{Sangjin Sin}. However, in this
section we will focus on the confinement construction of this solution. 

The most simple way to obtain a confinement theory is to follow the
discussion in \cite{Witten confinement,BBS book}. Specifically the
first step is to take one of the three spatial dimensions $x^{i}$
to be compactified on a circle with period $x^{i}\sim x^{i}+\delta x^{i}$.
So the dual theory ($\mathcal{N}=4$ Super Yang-Mills theory) becomes
effectively 3-dimensional (3d) below the Kaluza-Klein energy scale
$M_{KK}=1/\delta x^{i}$. Then the second step is to get rid of all
massless particles other than the gauge fields. The most convenient
way to achieve this is to require the fermion fields to be anti-periodic
on the circle while the bosons are given periodic boundary conditions.
Hence the supersymmetric fermions acquire mass of order $M_{KK}$
and the scalars of the SYM theory also get masses of order $M_{KK}$
induced by radiative corrections. Therefore the supersymmetry and
conformal symmetry are broken down and the resultant theory would
be three-dimensional YM theory below the energy scale $M_{KK}$. Next
we have to identify the bulk supergravity geometry as its holographic
correspondence. A trick for obtaining the answer is to perform a double
Wick rotation on the D(-1)-D3 brane background i.e. $t\rightarrow-ix^{i},x^{i}\rightarrow-it$.
Without loss of generality, let us denote $x^{i}=x^{3}$, then the
confining solution for D3-branes with smeared D-instantons reads,

\begin{align}
ds^{2} & =e^{\phi/2}\left\{ \frac{r^{2}}{R^{2}}\left[-dt^{2}+\left(dx^{1}\right)^{2}+\left(dx^{2}\right)^{2}+f\left(r\right)\left(dx^{3}\right)^{2}\right]+\frac{1}{f\left(r\right)}\frac{R^{2}}{r^{2}}dr^{2}+R^{2}d\Omega_{5}^{2}\right\} ,\nonumber \\
e^{\phi} & =1+\frac{Q}{r_{KK}^{4}}\log\frac{1}{f\left(r\right)},\ C=-e^{-\phi}+C_{0},\ f\left(r\right)=1-\frac{r_{KK}^{4}}{r^{4}},\ F_{5}=\mathcal{Q}_{3}\epsilon_{5}\label{eq:2}
\end{align}
where $C$ is the RR 0-form with $F_{1}=dC$, $R^{4}=4\pi g_{s}l_{s}^{4}$
and $Q,\mathcal{Q}_{3}$ relates to the charge of the D-instantons
and D3-branes. $\epsilon_{5}$ represents the volume form of a unit
$S^{5}$. The above solution is defined for $r>r_{KK}$ only thus
it does not have a horizon. This means $r=r_{KK}$ is the end of the
spacetime. Since the warp factor $e^{\phi/2}\frac{r^{2}}{R^{2}}$
never goes to zero, the asymptotics of the Wilson loop in this geometry
would lead to an area law which implies the holographically dual field
theory exhibits confinement below $M_{KK}$.  In order to avoid conical singularities in the
dual field theory, the following constraint has to be additionally
required,

\begin{equation}
M_{KK}=\frac{2r_{KK}}{R^{2}}.
\end{equation}
We note that if $Q\rightarrow0$ (i.e. no D-instantons) the supergravity
solution (\ref{eq:2}) consistently returns to the confining D3-brane
solution which is used in \cite{BBS book,Confining D3 - 1,Confining D3 - 2}.
The dual field theory can be examined by considering a probe D3-brane
located at $r\rightarrow\infty$ whose action is

\begin{align}
S_{\mathrm{D}3} & =\left[-T_{\mathrm{D}3}\int d^{4}xe^{-\frac{\phi}{2}}\mathrm{Str}\sqrt{-\det\left(g+\mathcal{F}\right)}+T_{\mathrm{D}3}\int C_{4}+\frac{1}{2}T_{\mathrm{D}3}\mathrm{Tr}\int C_{0}\mathcal{F}\land\mathcal{F}\right]\bigg|_{r\rightarrow\infty}\nonumber \\
 & \simeq-\frac{1}{4g_{4,YM}^{2}}\int d^{4}xF_{\mu\nu}F^{\mu\nu}+\frac{\kappa}{2}\mathrm{Tr}\int F\wedge F+\mathcal{O}\left(F^{3}\right)\nonumber \\
 & =-\frac{1}{4g_{3,YM}^{2}}\int d^{3}xF_{ab}F^{ab}+\frac{\kappa}{2}\mathrm{Tr}\int\omega_{3}.\label{eq:4}
\end{align}
Here$\mathcal{F}=2\pi\alpha^{\prime}F$ denotes the gauge field strength
and $T_{\mathrm{D3}}$ denotes the tension of D3-brane. $g_{4,3,YM}$
refers to the 4d and 3d Yang-Mills coupling constant respectively
and $\omega_{3}$ is the Chern-Simons (CS) 3-form. $\kappa$ corresponds
to the boundary value of $C$. Accordingly, we can conclude the dual
field theory to the background (\ref{eq:2}) is 3d confined YM plus
CS theory below $M_{KK}$ in holography.

\subsection{The confining D0-D4 solution}

Basically the confining D0-D4 solution where the D0-brane plays the
role of D-instanton can be achieved by following the same discussion
in the last section. Its confining solution reads \cite{WuChao D0-D4,Sangjin Sin D0-D4},

\begin{align}
ds^{2} & =\left(\frac{r}{R}\right)^{3/2}\left[H_{0}^{1/2}\eta_{\mu\nu}dx^{\mu}dx^{\nu}+H_{0}^{-1/2}f\left(r\right)\left(dx^{4}\right)^{2}\right]+H_{0}^{1/2}\left(\frac{R}{r}\right)^{3/2}\left[\frac{1}{f\left(r\right)}dr^{2}+r^{2}d\Omega_{4}^{2}\right],\nonumber \\
e^{\phi} & =\left(\frac{r}{R}\right)^{3/4}H_{0}^{3/4},\ f\left(r\right)=1-\frac{r_{KK}^{3}}{r^{3}},\ H_{0}=1+\frac{Q}{r^{3}},\nonumber \\
F_{2} & =dC_{1}=\frac{\mathcal{Q}_{0}}{r^{4}}\frac{1}{H_{0}^{2}}dr\wedge dx^{4},\ F_{4}=dC_{3}=\mathcal{Q}_{4}\epsilon_{4},\label{eq:5}
\end{align}
where $Q,\mathcal{Q}_{0}$ are two constants related to the number
density of D0-branes and $\mathcal{Q}_{4}$ is the charge of D4-branes.
We note that the D0-brane is the D-instanton in this system which
extends along the direction $x^{4}$ and $x^{4}$ is periodic $x^{4}\sim x^{4}+\delta x^{4}$.
To avoid conical singularities in the dual field theory, it leads
to the constraint

\begin{equation}
M_{KK}=\frac{3}{2}\frac{r_{KK}^{1/2}}{R^{3/2}}\frac{1}{\sqrt{1+Q/r_{KK}^{3}}}.
\end{equation}
Below $M_{KK}$ the dual field theory is confined theory which can
be investigated by introducing a probe D4-brane in the background
(\ref{eq:5}) at $r\rightarrow\infty$. Then its low-energy action
is

\begin{align}
S_{\mathrm{D}4} & =\left[-T_{\mathrm{D}4}\int d^{5}xe^{-\phi}\mathrm{Str}\sqrt{-\det\left(g+\mathcal{F}\right)}+T_{4}\int C_{5}+\frac{1}{2}T_{\mathrm{D}4}\mathrm{Tr}\int C_{1}\land\mathcal{F}\land\mathcal{F}\right]\bigg|_{r\rightarrow\infty}\nonumber \\
 & \simeq-\frac{1}{4g_{4,YM}^{2}}\int d^{4}xF_{\mu\nu}F^{\mu\nu}+\frac{\theta}{16\pi^{2}}\mathrm{Tr}\int F\wedge F+\mathcal{O}\left(F^{3}\right).
\end{align}
We note that the theta angle corresponds to $\theta\sim\int_{S_{x^{4}}}C_{1}$.
Therefore the dual theory in the D0-D4 system is theta-dependent confined
Yang-Mills theory \cite{WuChao D0-D4,Francesco D0-D4 - 1,Francesco D0-D4 - 2}.

\section{Potential analysis }

In this section, let us perform the analysis by following \cite{Y. Sato,Confining D3 - 1,Confining D3 - 2}
for the confining background with D-instantons to evaluate the effective
potential in Schwinger effect.

\subsection{The D(-1)-D3 background}

\begin{figure}
\noindent \begin{centering}
\includegraphics[scale=0.2]{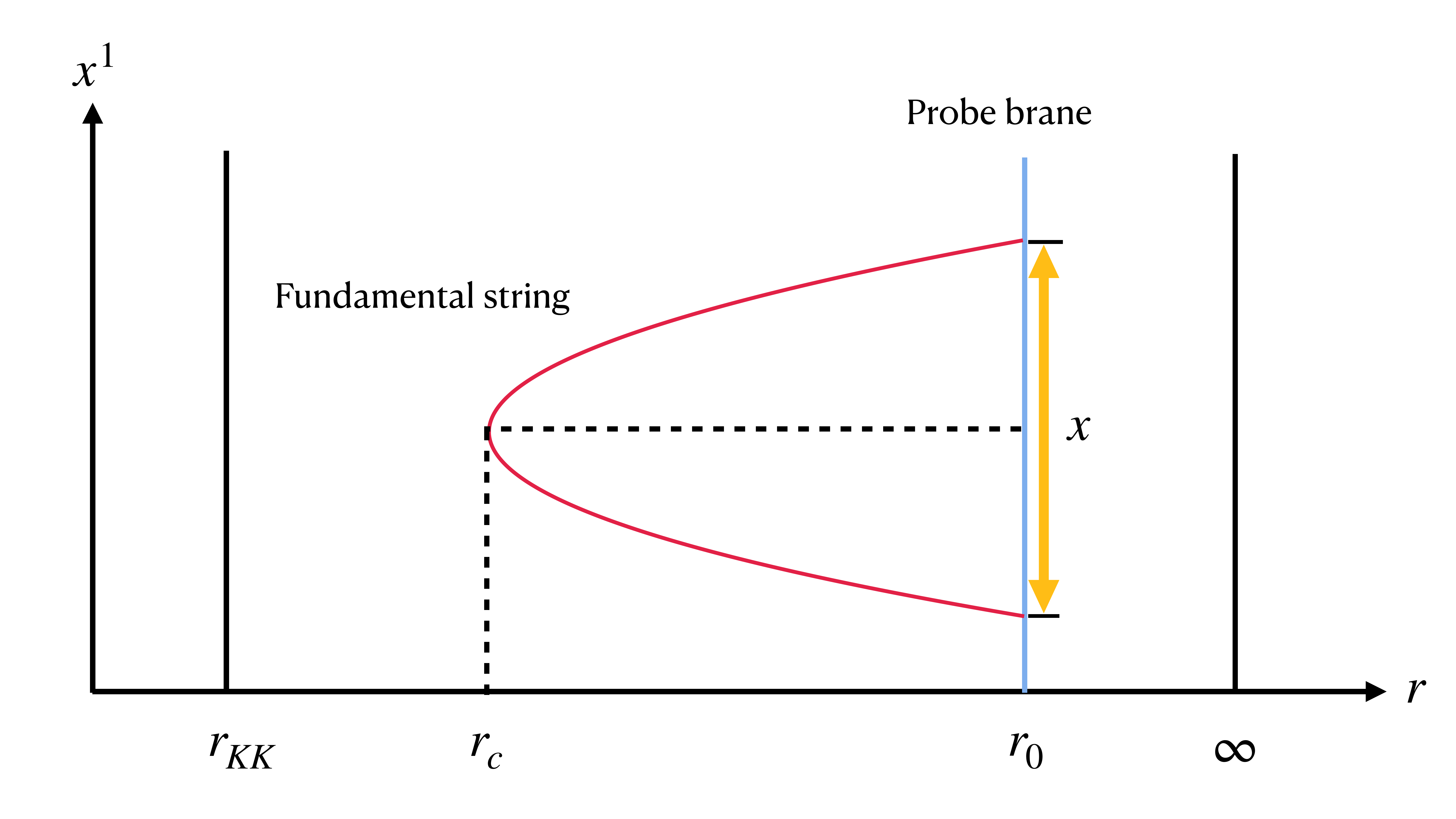}
\par\end{centering}
\caption{\label{fig:1}Configuration of the string and probe brane.}

\end{figure}

\begin{figure}
\begin{centering}
\includegraphics[scale=0.35]{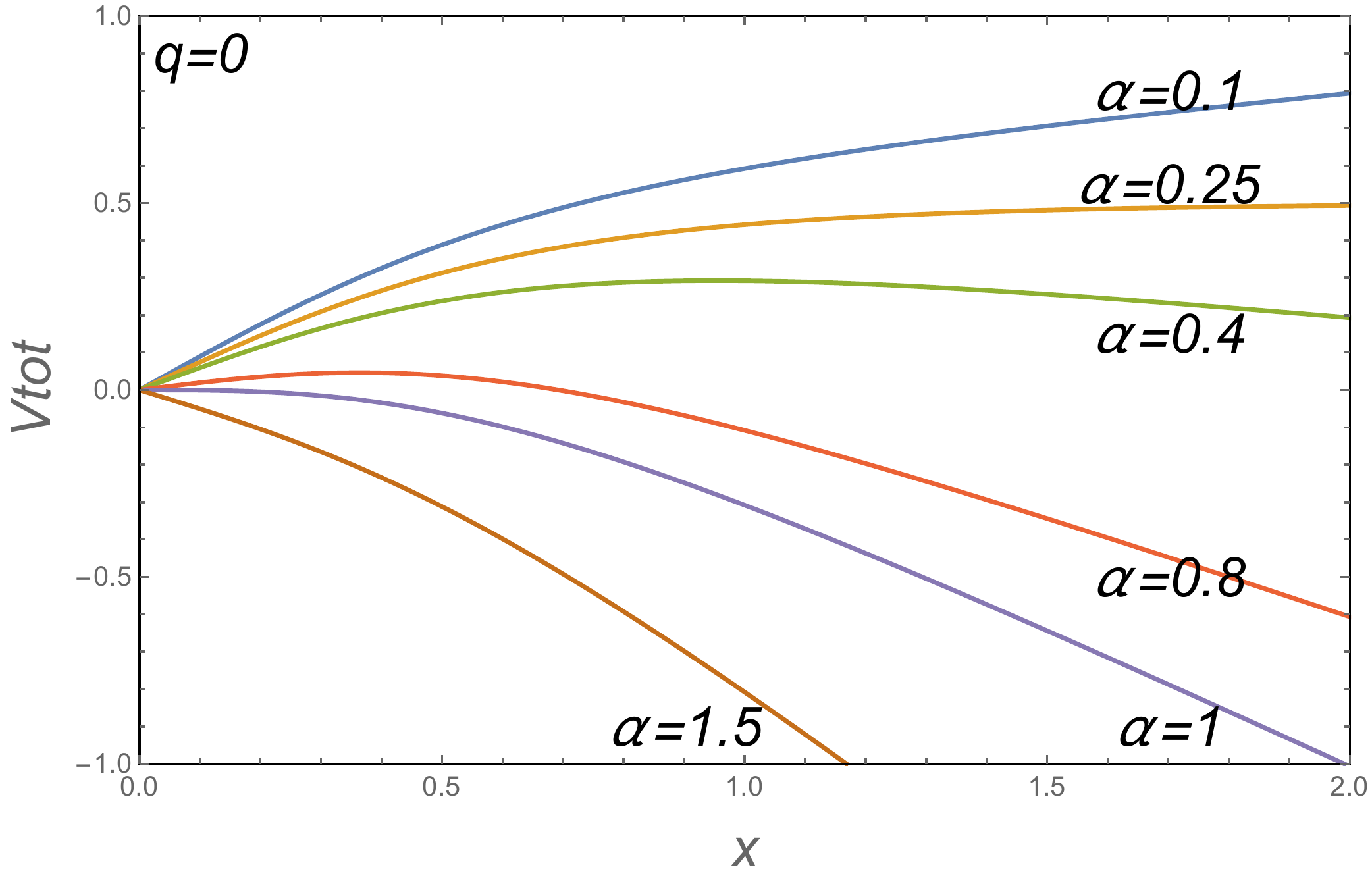}\includegraphics[scale=0.35]{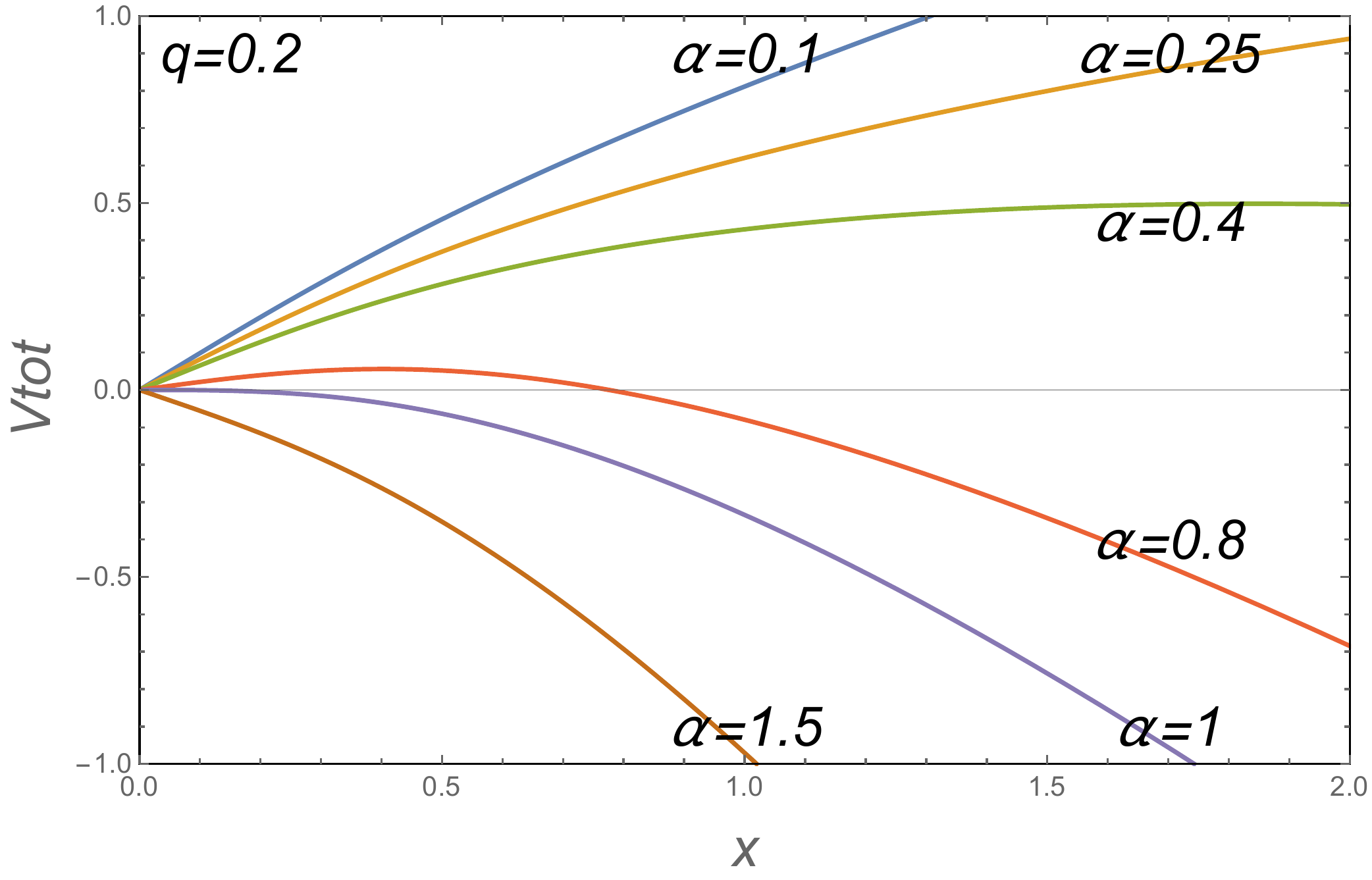}
\par\end{centering}
\begin{centering}
\includegraphics[scale=0.35]{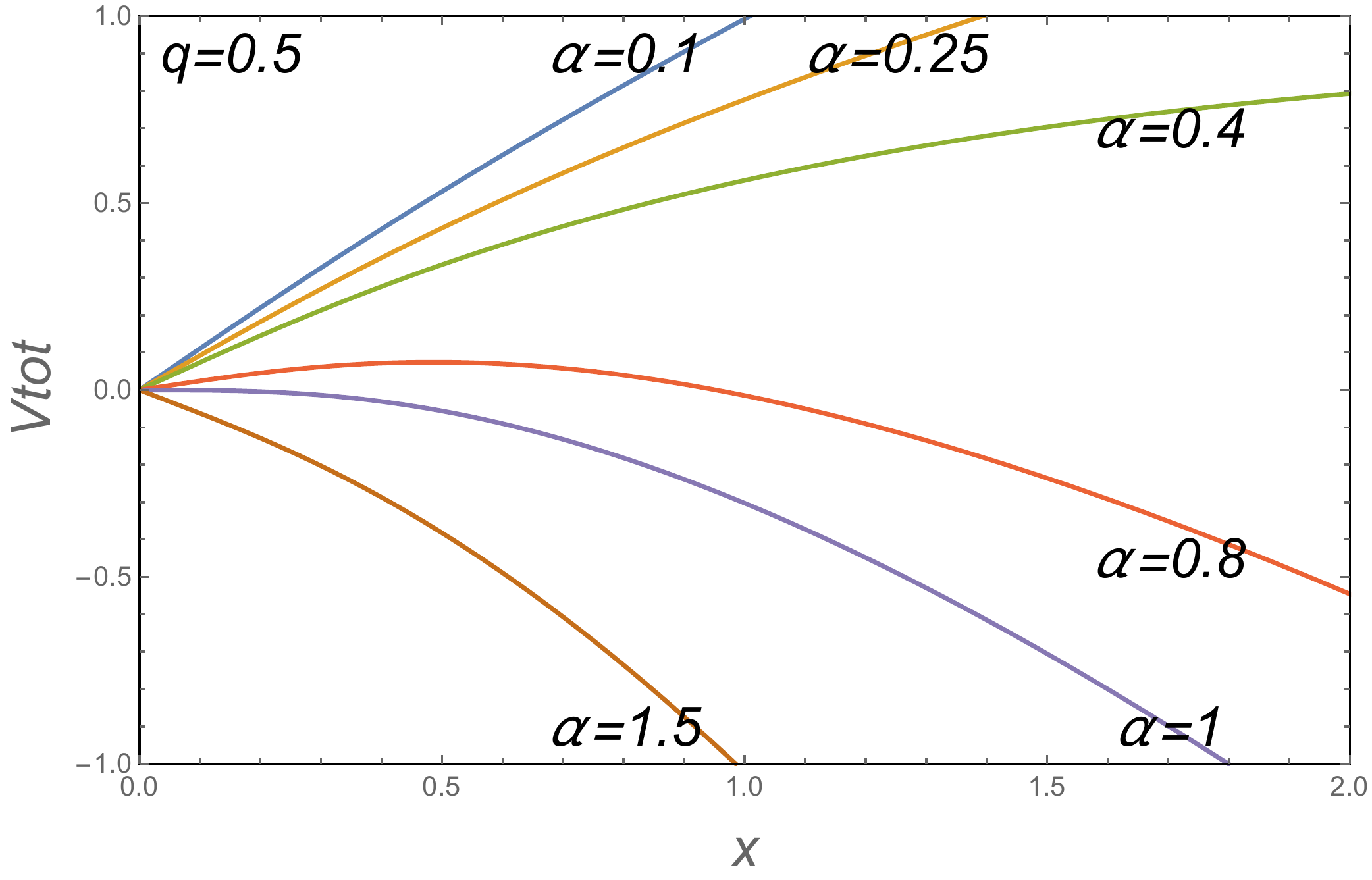}\includegraphics[scale=0.35]{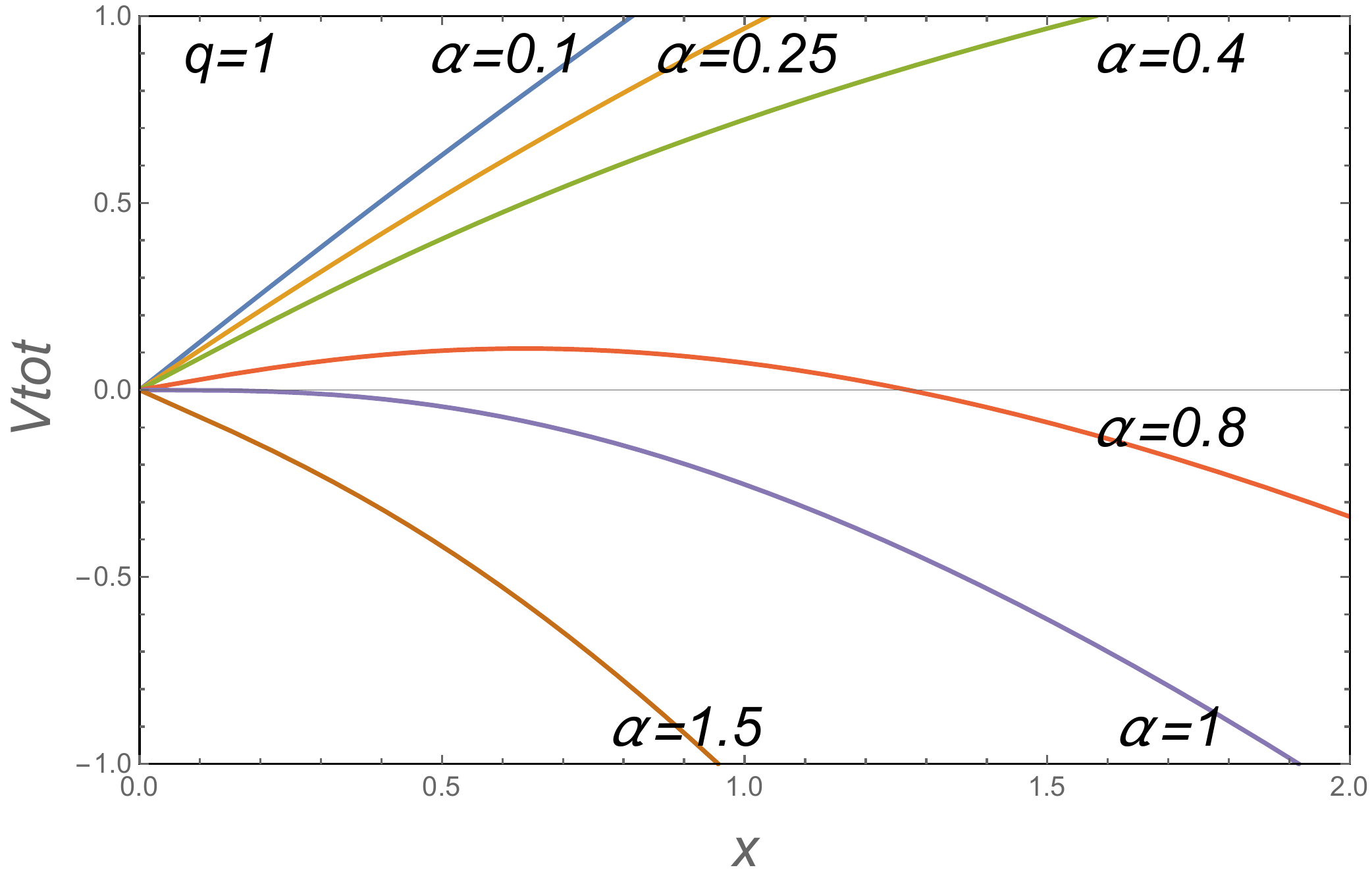}
\par\end{centering}
\caption{\label{fig:2}The dependence on $x$ of the total potential $V_{tot}\left(x\right)$
with various instanton densities and electric fields (determined by
$\alpha$) in the confining D(-1)-D3 background. In all graphs we
have set $b=0.5$ and $2L^{2}/r_{0}=2T_{f}r_{0}=1$.}

\end{figure}

\begin{figure}
\begin{centering}
\includegraphics[scale=0.45]{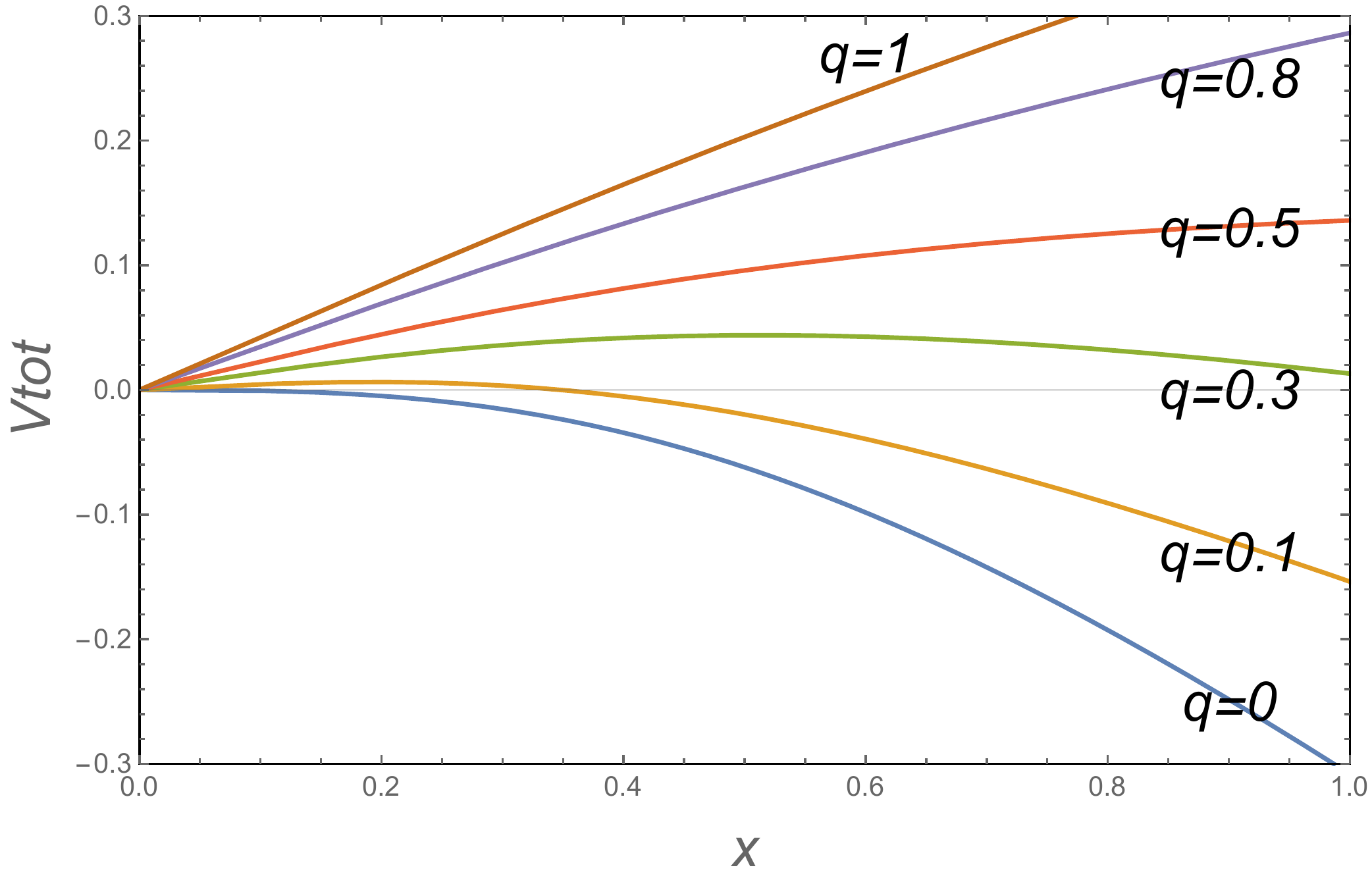}
\par\end{centering}
\caption{\label{fig:3} The relation of $V_{tot}\left(x\right)$ and $q$ in
the D(-1)-D3 background with $b=0.5,\tilde{\alpha}=1$.}
\end{figure}

In order to study the Schwinger effect, we need to evaluate the effective
potential and find the critical value of the electric field first.
So let us start with a probe D3-brane located at $r=r_{0}$ on which
an external electric field $F_{01}=E$ is switched. The DBI (Dirac-Born-Infeld)
action of the probe brane is given as,

\begin{align}
S & =-T_{\mathrm{D}3}\int d^{4}xe^{-\frac{\phi}{2}}\sqrt{-\det\left(g+\mathcal{F}\right)}\nonumber \\
 & =-T_{\mathrm{D}3}V_{4}\frac{r_{0}^{4}}{R^{4}}e^{\frac{\phi\left(r_{0}\right)}{2}}f\left(r_{0}\right)^{1/2}\sqrt{1-\frac{\left(2\pi\alpha^{\prime}\right)^{2}R^{4}}{e^{\phi\left(r_{0}\right)}r_{0}^{4}}E^{2}}.
\end{align}
It is easy to understand that the classical solution would not exist
if $E>E_{c}$ where $E_{c}$ is a critical value of the electric field.
Thus the critical electric field can be obtained as, 

\begin{equation}
E_{c}=\frac{1}{2\pi\alpha^{\prime}}\frac{r_{0}^{2}}{R^{2}}e^{\frac{\phi\left(r_{0}\right)}{2}}.\label{eq:9}
\end{equation}
Then we need to calculate the total energy of a pair of the fundamental
particles which can be computed by evaluating the expectation of its
rectangular Wilson loop. In holography it corresponds to the world-sheet
area or equivalently the on-shell Nambu-Goto (NG) action of a fundamental
string \cite{Maldacena Wilson loop}. Accordingly we consider a Fermion-anti
Fermion pair placed at fixed positions on the probe brane with separation
$x$ in direction $x^{1}$. Choosing the static gauge, the induced
metric on the world sheet with $\tau=t,x^{1}=\sigma,r=r\left(\sigma\right)$
is,

\begin{equation}
ds^{2}=g_{\alpha\beta}dx^{\alpha}dx^{\beta}=\frac{r^{2}}{R^{2}}e^{\phi/2}\left\{ d\tau^{2}+\left[1+\frac{1}{f\left(r\right)}\frac{R^{4}}{r^{4}}\left(\frac{dr}{d\sigma}\right)^{2}\right]\left(d\sigma\right)^{2}\right\} ,\ \ \ \ \alpha,\beta=0,1
\end{equation}
where we have worked in the Euclidean signature. Therefore the NG
action is calculated as,

\begin{align}
S_{NG} & =T_{f}\int d\tau d\sigma\sqrt{\det\left(g_{\alpha\beta}\right)}\nonumber \\
 & =T_{f}\int d\tau d\sigma\sqrt{e^{\phi\left(r\right)}\left[\frac{1}{f\left(r\right)}\left(\frac{dr}{d\sigma}\right)^{2}+\frac{r^{4}}{R^{4}}\right]}.\label{eq:11}
\end{align}
Since the Lagrangian in (\ref{eq:11}) does not depend on $\sigma$
explicitly, its associated Hamiltonian is conserved i.e. a constant,

\begin{equation}
\mathcal{H}=\left(\partial_{\sigma}r\right)\frac{\partial\mathcal{L}}{\partial_{\sigma}r}-\mathcal{L}=const,
\end{equation}
which means

\begin{equation}
\frac{e^{\frac{\phi\left(r\right)}{2}}r^{4}/R^{4}}{\sqrt{\frac{r^{4}}{R^{4}}+\frac{1}{f\left(r\right)}\left(\frac{dr}{d\sigma}\right)^{2}}}=const.=e^{\frac{\phi\left(r_{c}\right)}{2}}\frac{r_{c}^{2}}{R^{2}},\label{eq:13}
\end{equation}
if the boundary condition

\begin{equation}
\frac{dr}{d\sigma}\bigg|_{r=r_{c}}=0,\ \sigma=\sigma_{0}
\end{equation}
is imposed where $r_{c}$ refers to the top point of the string in
the bulk as illustrated in Figure \ref{fig:1}.
 So (\ref{eq:13}) leads to 

\begin{equation}
\frac{dr}{d\sigma}=\frac{1}{R^{2}}\sqrt{\left[e^{\phi\left(r\right)-\phi\left(r_{c}\right)}\frac{r^{4}}{r_{c}^{4}}-1\right]\left(r^{4}-r_{KK}^{4}\right)},
\end{equation}
and the separation $x$ is therefore obtained as

\begin{equation}
x=\frac{2R^{2}}{r_{0}a}e^{\phi\left(1\right)/2}\int_{1}^{1/a}\frac{dy}{\sqrt{\left(y^{4}-b^{4}/a^{4}\right)\left[e^{\phi\left(y\right)-\phi\left(1\right)}y^{4}-1\right]}},
\end{equation}
where we have used the dimensionless quantities defined as,

\begin{equation}
y=\frac{r}{r_{c}},\ a=\frac{r_{c}}{r_{0}},\ b=\frac{r_{KK}}{r_{0}},\ \frac{Q}{r_{0}^{4}}=q.
\end{equation}
Afterwards the potential energy (PE) including static energy (SE)
is computed as

\begin{equation}
V_{\mathrm{PE+SE}}=2T_{f}\int_{0}^{x/2}d\sigma\mathcal{L}=2T_{f}r_{0}a\int_{1}^{1/a}\frac{e^{\phi\left(y\right)}y^{4}}{\sqrt{\left(y^{4}-b^{4}/a^{4}\right)\left[e^{\phi\left(y\right)-\phi\left(1\right)}y^{4}-1\right]}}.
\end{equation}
However we have to add the contribution from the interaction of the
Fermion-anti Fermion pair with an external electric field $E$ to
this energy. So the total potential is,

\begin{align}
V_{\mathrm{tot}}\left(x\right) & =V_{\mathrm{PE+SE}}-Ex\nonumber \\
 & =2T_{f}r_{0}a\int_{1}^{1/a}\frac{e^{\phi\left(y\right)}y^{4}}{\sqrt{\left(y^{4}-b^{4}/a^{4}\right)\left[e^{\phi\left(y\right)-\phi\left(1\right)}y^{4}-1\right]}}\nonumber \\
 & -\frac{2T_{f}r_{0}\alpha}{a}e^{\phi\left(1\right)}\int_{1}^{1/a}\frac{dy}{\sqrt{\left(y^{4}-b^{4}/a^{4}\right)\left[e^{\phi\left(y\right)-\phi\left(1\right)}y^{4}-1\right]}},
\end{align}
where

\begin{equation}
\alpha=\frac{E}{E_{c}\left(q\right)},\ E_{c}\left(q\right)=T_{f}\frac{r_{0}^{2}}{R^{2}}e^{\phi\left(1\right)/2},
\end{equation}
and $E_{c}$ is the critical electric field given in (\ref{eq:9}).
The relation of $V_{tot}$ and $x$ can be evaluated numerically which
has been illustrated as in Figure \ref{fig:2} with various $q,\alpha$.
 In Figure \ref{fig:2} we have set $b=0.5$ (fixing energy scale
$M_{KK}$) , $\alpha=0.4,0.6,0.8,1,1.2$ and the instanton density
has been chosen as $q=0,0.2,0.5,1$ respectively in the four panels
for comparison. These graphs imply that the potential barrier vanishes
for $\alpha>1$, so the critical electric field obtained from the
potential analysis agrees with (\ref{eq:9}) which is evaluated by
using the DBI action as expected in \cite{Confining D3 - 1,general SE}.
Besides we notice that the presence of instantons increases the potential
barrier and thus suppresses the pair creation. This conclusion would
be further confirmed by analyzing the formula (\ref{eq:11}) of the
NG action and the expectation of a circular Wilson loop in the following
sections. Nonetheless, to see this quickly, let us introduce another
dimensionless parameter $\tilde{\alpha}=\frac{E}{E_{c}\left(q=0\right)}$
so that $E=E_{c}\left(q=0\right)$ corresponds to $\tilde{\alpha}=1$,
then the numerical result of $V_{tot}\left(x\right)$ and $q$ is
shown in Figure \ref{fig:3}. 
 We find there is not any limitation to pair creation at zero instanton
density since there is no potential barrier in the presence of electric
field for $q=0$. As we can see the presence of the instantons develops
a potential barrier, so the Schwinger effect occurs through a tunneling
process only if $q>0$ and the instanton density increases the potential
barrier whereas suppresses the pair creation. It therefore implies
that the critical electric field is also increased by the instanton
density which is in agreement with (\ref{eq:9}). And all these conclusions
agree with the results evaluated in the deconfined D(-1)-D3 background
at zero temperature limit in \cite{deconfined D(-1)-D3}.

\subsection{The D0-D4 background}

In this section, let us turn to the D0-D4 case. The discussion would
be basically parallel to those in Section 3.2. First we consider the
DBI action of a probe D4-brane in the D0-D4 background with an electric
field located at $r=r_{0}$. The action takes the form,

\begin{align}
S & =-T_{\mathrm{D4}}\int d^{5}xe^{-\phi}\sqrt{-\det\left(g+\mathcal{F}\right)}\nonumber \\
 & =-T_{\mathrm{D4}}V_{5}\frac{r^{3}}{R^{3}}f\left(r_{0}\right)^{1/2}\sqrt{1-\frac{\left(2\pi\alpha^{\prime}\right)^{2}R^{3}}{H_{0}\left(r_{0}\right)r_{0}^{3}}E^{2}}.
\end{align}
Hence the critical electric field is evaluated as,

\begin{equation}
E_{c}=\frac{1}{2\pi\alpha^{\prime}}\frac{r_{0}^{3/2}}{R^{3/2}}H_{0}^{1/2}\left(r_{0}\right).\label{eq:22}
\end{equation}
Then the Euclidean induced metric on the world sheet of a fundamental
string with the choice of static gauge $\tau=t,x^{1}=\sigma,r=r\left(\sigma\right)$
is given as,

\begin{equation}
ds^{2}=\left(\frac{r}{R}\right)^{3/2}H_{0}^{1/2}\left\{ d\tau^{2}+\left[1+\frac{1}{f\left(r\right)}\left(\frac{R}{r}\right)^{3}\left(\frac{dr}{d\sigma}\right)^{2}\right]d\sigma^{2}\right\} .
\end{equation}
\begin{figure}[h]
\begin{centering}
\includegraphics[scale=0.35]{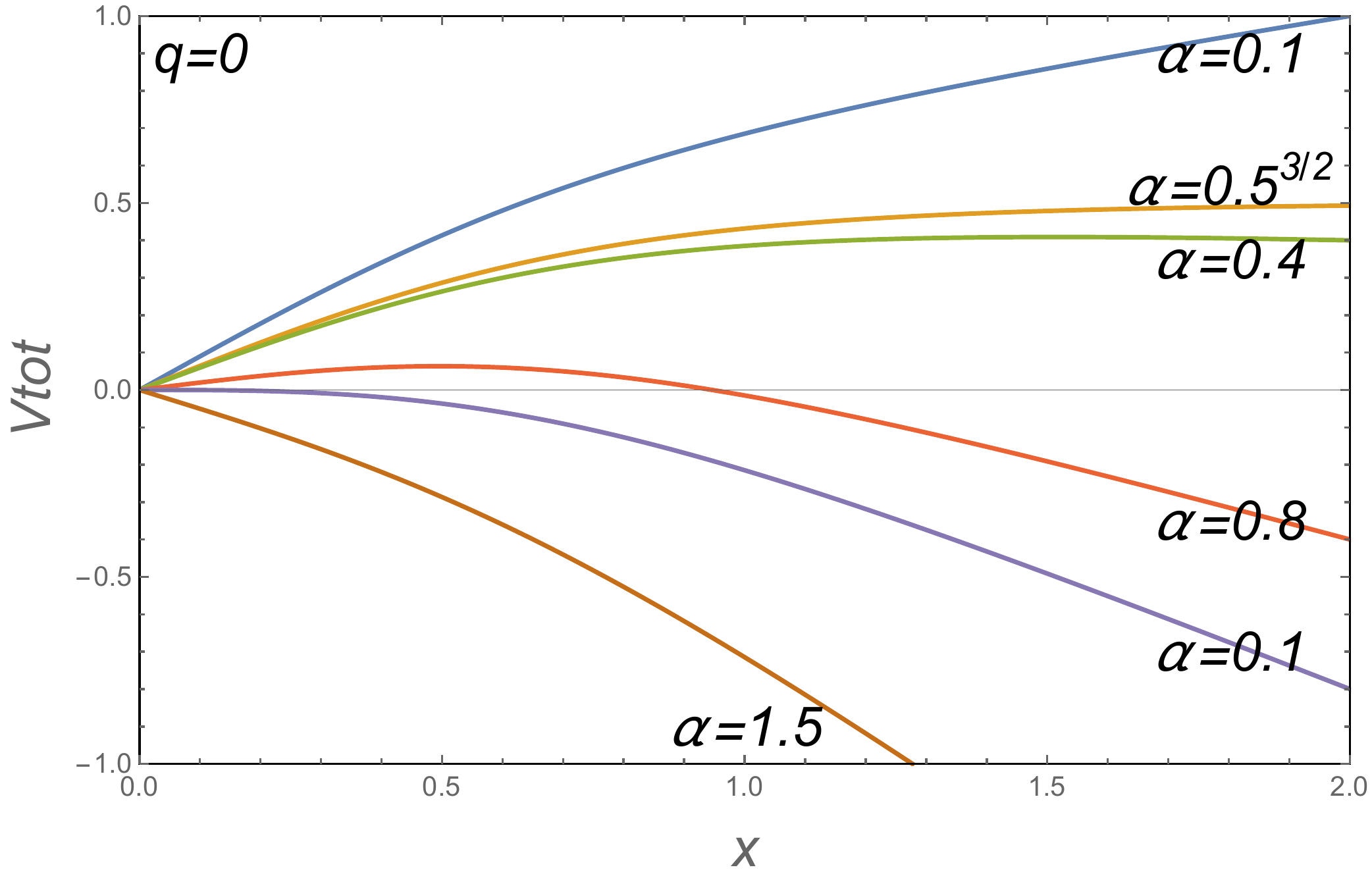}\includegraphics[scale=0.35]{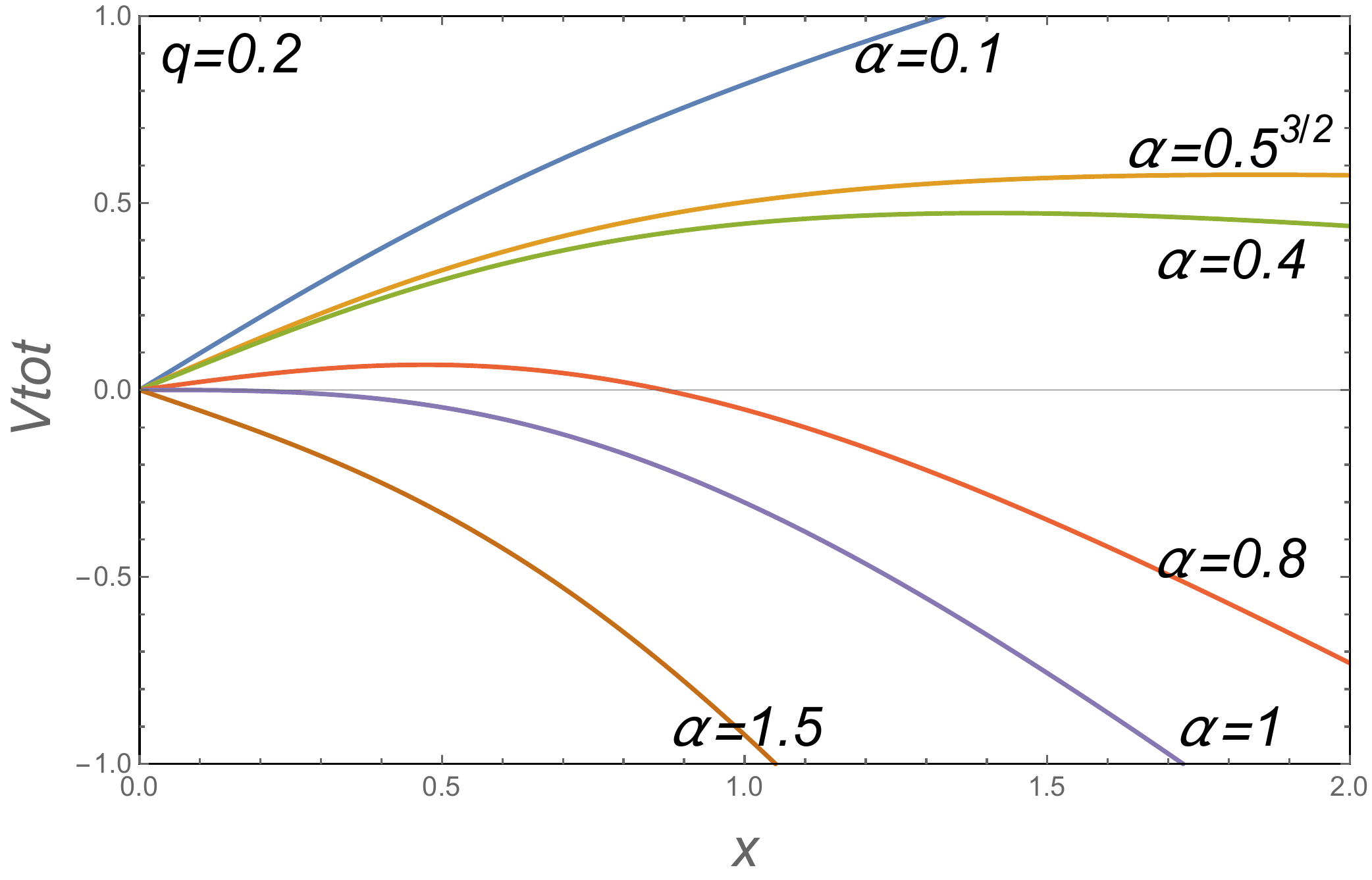}
\par\end{centering}
\begin{centering}
\includegraphics[scale=0.35]{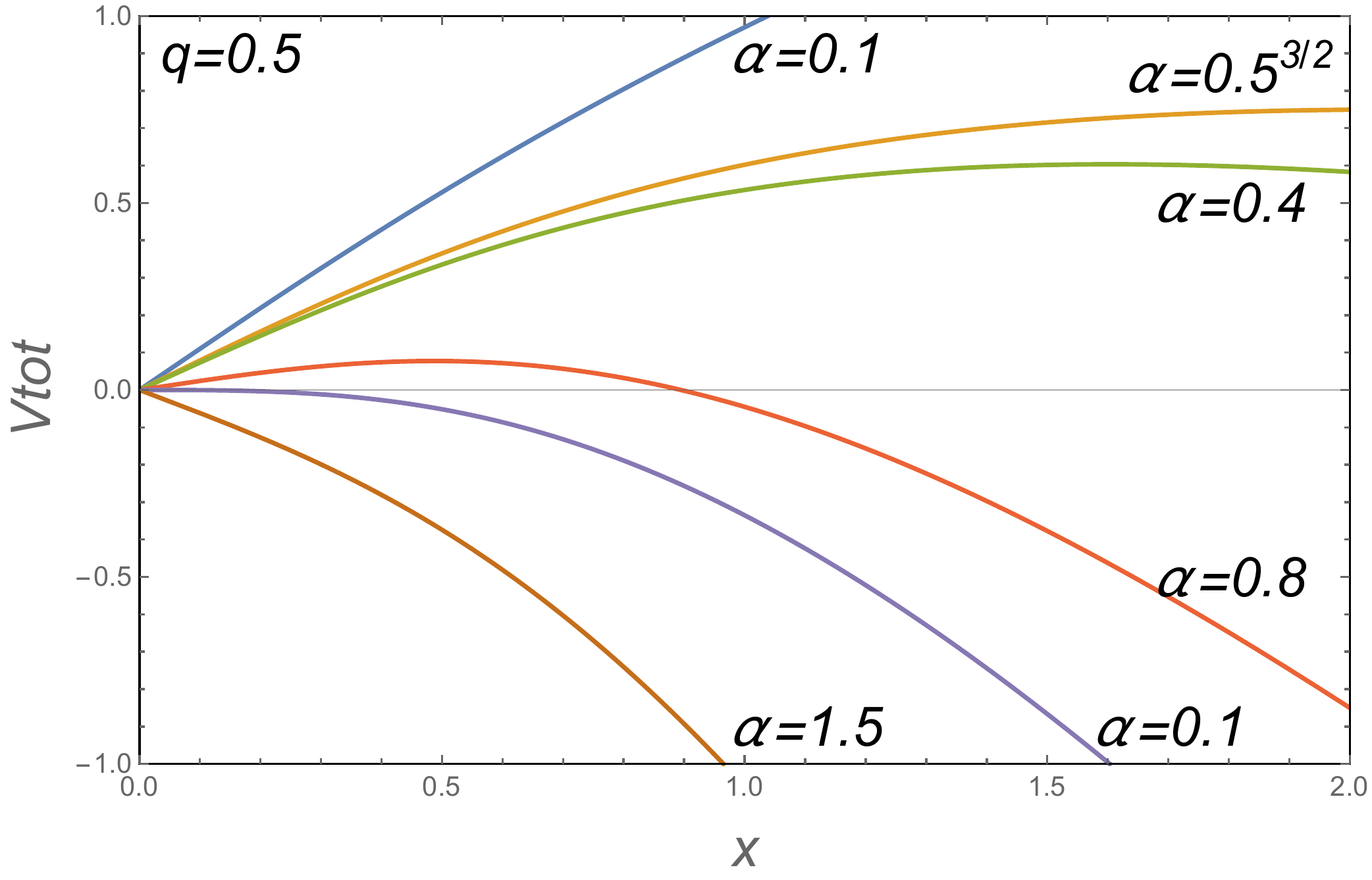}\includegraphics[scale=0.35]{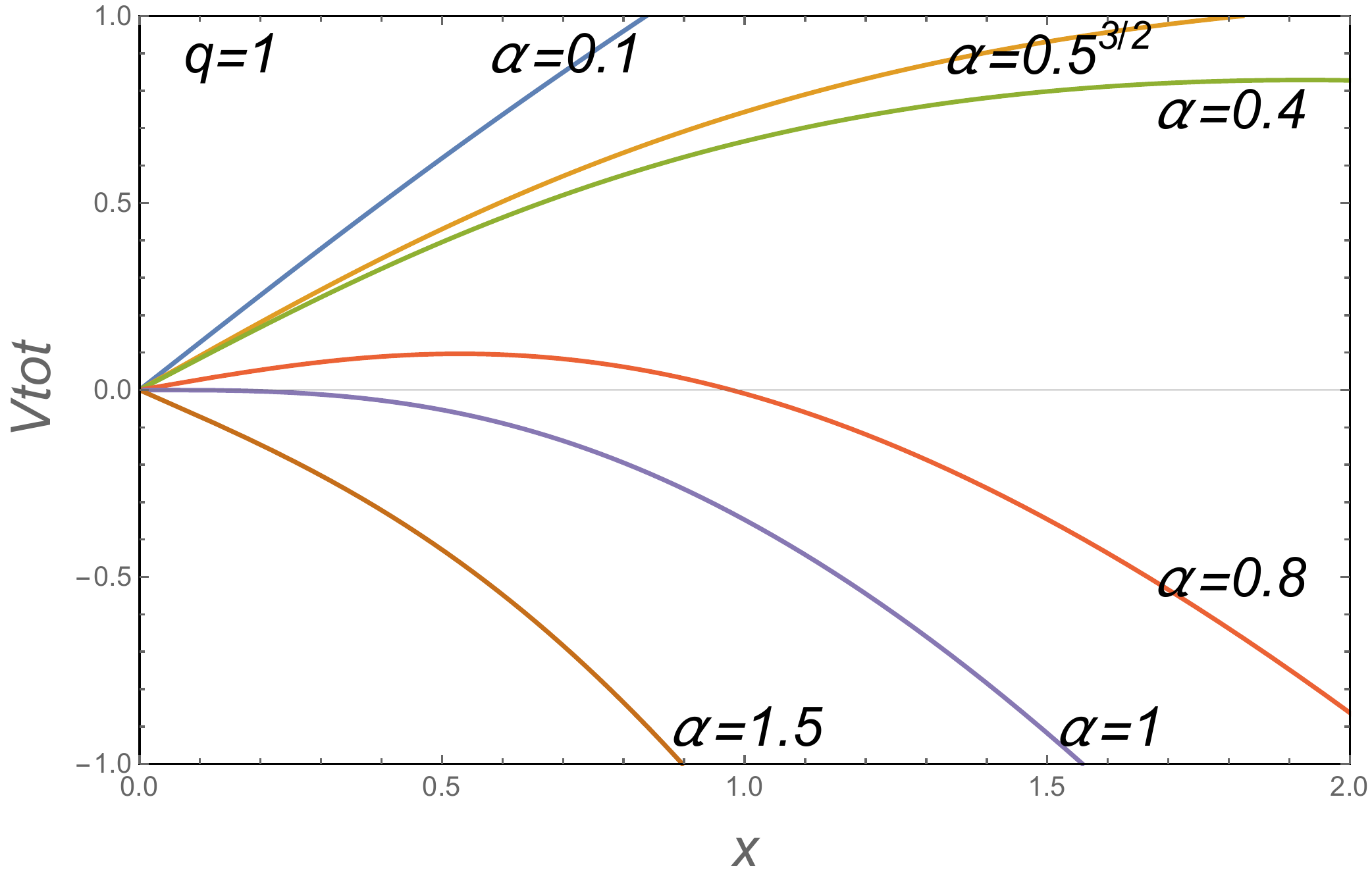}
\par\end{centering}
\caption{\label{fig:4} The dependence on $x$ of the total potential $V_{tot}\left(x\right)$
with various instanton densities and electric fields (determined by
$\alpha$) in the confining D0-D4 background. In all graphs we have
set $b=0.5$ and $2L^{3/2}/r_{0}^{1/2}=2T_{f}r_{0}=1$.}
\end{figure}
 Therefore the NG action can be computed as,

\begin{equation}
S_{NG}=T_{f}\int d\tau d\sigma H_{0}^{1/2}\left(r\right)\sqrt{\frac{r^{3}}{R^{3}}+\frac{1}{f\left(r\right)}\left(\frac{dr}{d\sigma}\right)^{2}.}
\end{equation}
The conserved Hamiltonian is

\begin{equation}
\frac{H_{0}^{1/2}\left(r\right)r^{3}/R^{3}}{\sqrt{\frac{r^{3}}{R^{3}}+\frac{1}{f\left(r\right)}\left(\frac{dr}{d\sigma}\right)^{2}}}=const.=\frac{r_{c}^{3/2}}{R^{3/2}}H_{0}^{1/2}\left(r_{c}\right),
\end{equation}
where we have used the boundary condition
\begin{equation}
\frac{dr}{d\sigma}\bigg|_{r=r_{c}}=0,\ \sigma=\sigma_{0}.
\end{equation}
So we have,

\begin{equation}
\frac{dr}{d\sigma}=\frac{1}{R^{3/2}H_{0}^{1/2}\left(r_{c}\right)}\sqrt{\left(r^{3}-r_{KK}^{3}\right)\left(\frac{r^{3}}{r_{c}^{3}}-1\right)},
\end{equation}
which leads to the formula of the separation $x$ as,

\begin{equation}
x=2\int_{r_{c}}^{r_{0}}dr=\frac{2R^{3/2}}{r_{0}^{1/2}a^{1/2}}H_{0}^{1/2}\left(1\right)\int_{1}^{1/a}\frac{dy}{\sqrt{\left(y^{3}-1\right)\left(y^{3}-b^{3}/a^{3}\right)}},
\end{equation}
with the dimensionless quantities introduced as,

\begin{equation}
y=\frac{r}{r_{c}},\ a=\frac{r_{c}}{r_{0}},\ b=\frac{r_{KK}}{r_{0}},\ \frac{Q}{r_{0}^{3}}=q.
\end{equation}
Afterwards the potential energy (PE) including static energy (SE)
in D0-D4 background is computed as,

\begin{equation}
V_{\mathrm{PE+SE}}=2T_{f}\int_{0}^{x/2}d\sigma\mathcal{L}=2T_{f}r_{0}a\int_{1}^{1/a}dy\frac{y^{3}H_{0}\left(y\right)}{\sqrt{\left(y^{3}-1\right)\left(y^{3}-b^{3}/a^{3}\right)}}.
\end{equation}
\begin{figure}[h]
\begin{centering}
\includegraphics[scale=0.45]{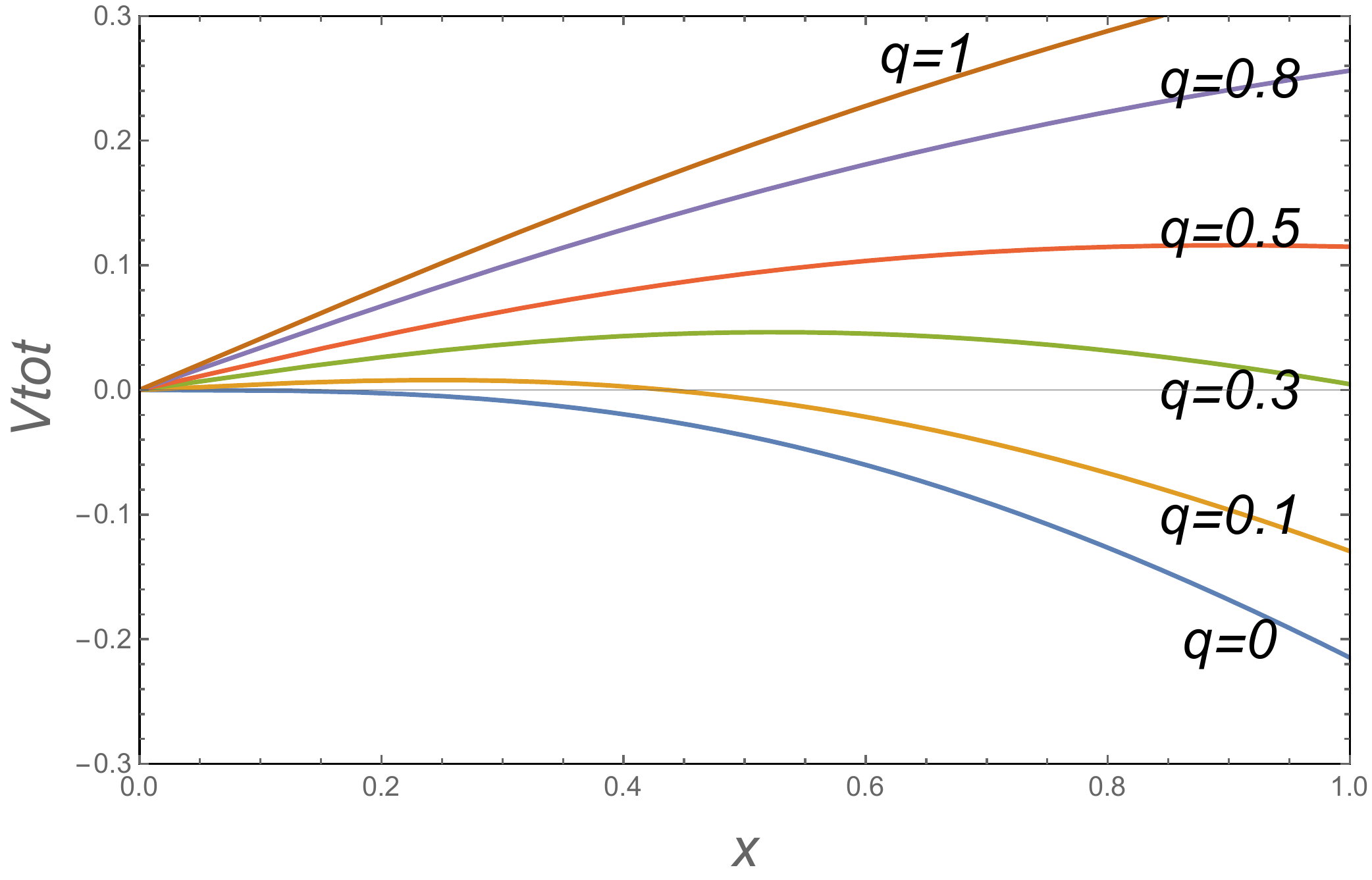}
\par\end{centering}
\caption{\label{fig:5} The relation of $V_{tot}\left(x\right)$ and $q$ in
the D(-1)-D3 background with $b=0.5^{3/2},\tilde{\alpha}=1$.}
\end{figure}
 By taking into account the contribution of the electric field, the
total potential can be obtained as,

\begin{align}
V_{\mathrm{tot}} & =V_{\mathrm{PE+SE}}-Ex\nonumber \\
 & =2T_{f}r_{0}a\int_{1}^{1/a}dy\frac{y^{3}H_{0}\left(y\right)}{\sqrt{\left(y^{3}-1\right)\left(y^{3}-b^{3}/a^{3}\right)}}-\frac{2T_{f}r_{0}\alpha}{a^{1/2}}H_{0}\left(1\right)\int_{1}^{1/a}\frac{dy}{\sqrt{\left(y^{3}-1\right)\left(y^{3}-b^{3}/a^{3}\right)}},
\end{align}
where

\begin{equation}
\alpha=\frac{E}{E_{c}},\ E_{c}=\frac{1}{2\pi\alpha^{\prime}}\frac{r_{0}^{3/2}}{R^{3/2}}H_{0}^{1/2}\left(r_{0}\right).
\end{equation}
The relation of $V_{tot}$ and $x$ in the D4-D0 background can be
evaluated numerically and the results are illustrated as in Figure
\ref{fig:4}. These graphs also show that the potential barrier vanishes
for $\alpha>1$ which agrees with our (\ref{eq:22}) and again the
potential analysis as expected in \cite{Confining D3 - 1,general SE}.
By introducing $\tilde{\alpha}=\frac{E}{E_{c}\left(q=0\right)}$,
we find the presence of the instantons develops a potential barrier
and increases it as illustrated in Figure \ref{fig:5}. So the Schwinger
effect can occur without any any limitation at $q=0$ while it happens
only through a tunneling process for $q>0$. Since the instantons
increase the potential barrier, it suppresses the pair creation. And
this conclusion is in agreement with the analysis from the flavor
brane approach for the Schwinger effect in the D0-D4 background in
\cite{Wenhe}.

\section{Pair production rate}

To exactly analyze the Schwinger effect, we are going to compute the
pair production rate in this section. As mentioned before, this quantity
can be obtained by evaluating the expectation of a circular Wilson
loop living in the $t-x$ plane in the presence of an external electric
field. Following the discussion in \cite{Confining D3 - 2}, the expectation
of the Wilson loop corresponds to the Euclidean version of the string
onshenll action in holography. So let us work in the Euclidean signature
and choose the polar coordinates in the $t-x$ plane as,

\begin{equation}
t=\rho\cos\eta,\sigma=\rho\sin\eta,\label{eq:33}
\end{equation}
with all constant other coordinates. Since the D(-1)-D3 and D0-D4
background are the concerns, we will analyze the string action respectively
in the two backgrounds.

\subsection*{The D(-1)-D3 case}

\begin{figure}
\begin{centering}
\includegraphics[scale=0.38]{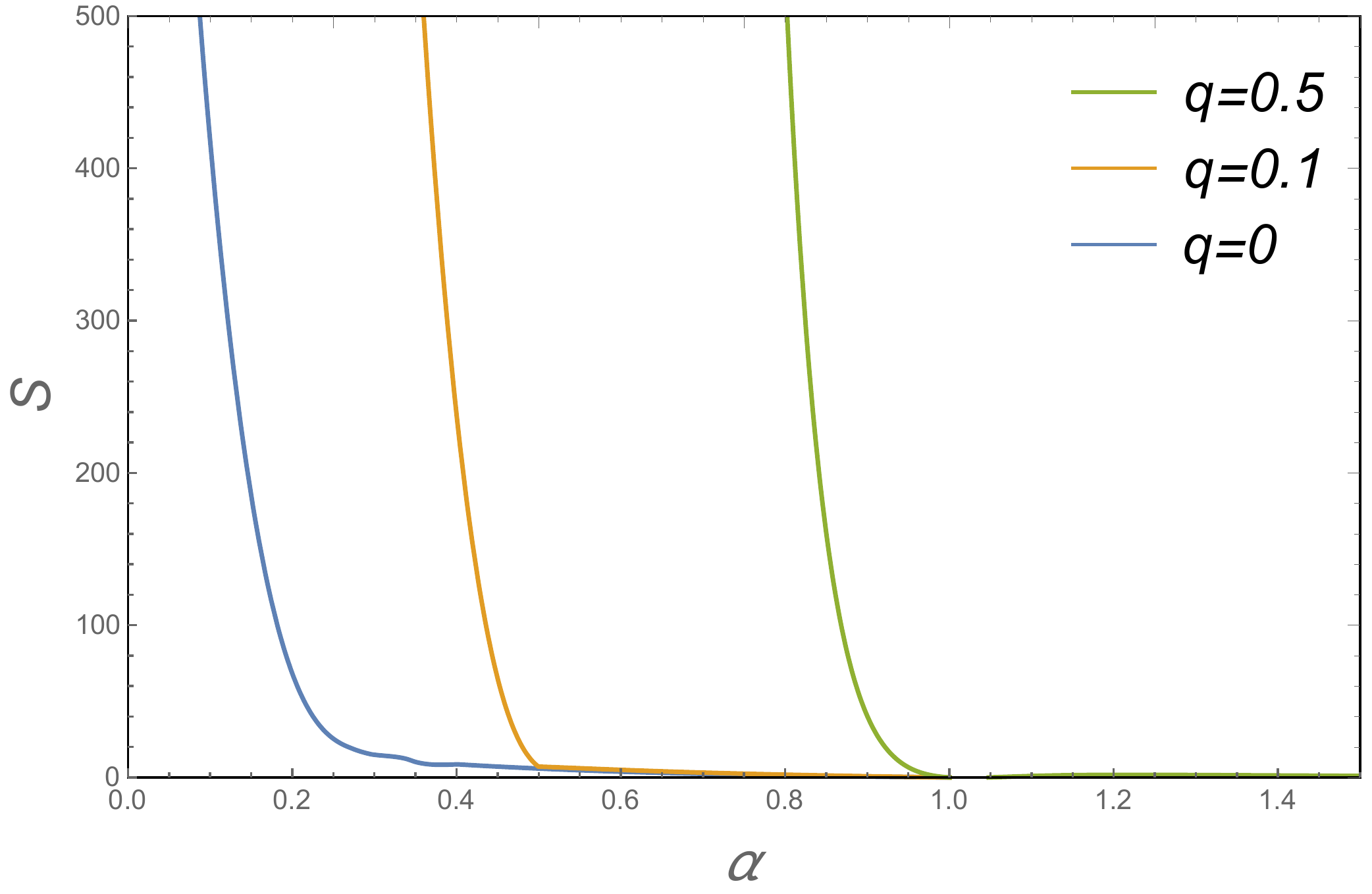}\includegraphics[scale=0.38]{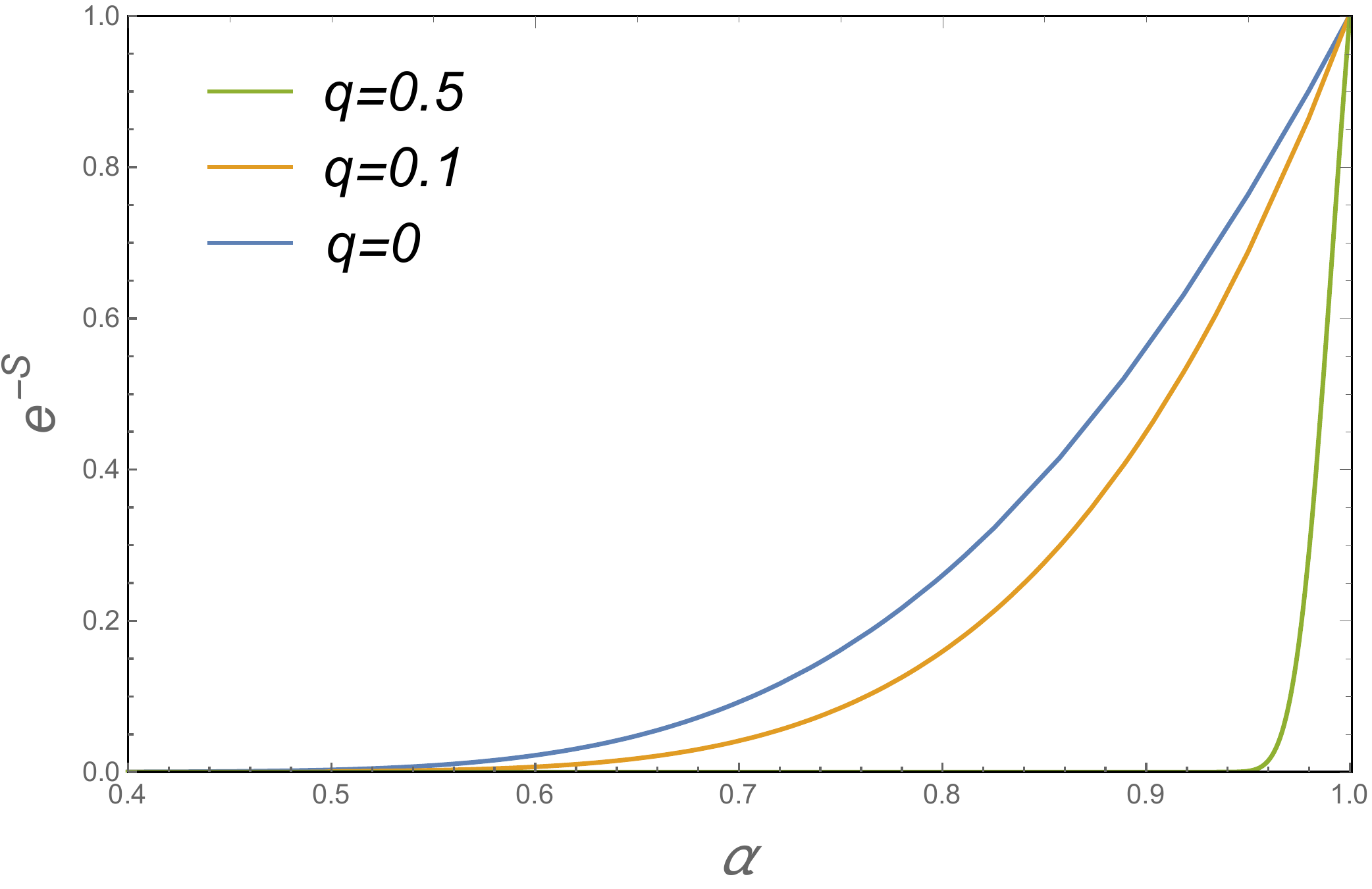}
\par\end{centering}
\caption{\label{fig:6} The classical string action and its exponential behavior
versus $\alpha$ in the D(-1)-D3 background.}

\end{figure}

Using (\ref{eq:11}) and (\ref{eq:33}), the string action can be
written as,

\begin{equation}
S=S_{NG}+S_{B_{2}},
\end{equation}
where

\begin{align}
S_{NG} & =2\pi T_{f}R^{2}\int d\rho e^{\phi/2}\frac{\rho}{z^{2}}\sqrt{1+\frac{z^{\prime2}}{f}},\ f=1-\frac{z^{4}}{z_{KK}^{4}},\nonumber \\
S_{B_{2}} & =-2\pi T_{f}B_{01}\int_{0}^{x}d\rho\rho=-\pi Ex^{2}.\label{eq:35}
\end{align}
Here we have defined the $z$ coordinate as $z\left(\rho\right)=R^{2}/r$
and the derivatives are with respect to $\rho$. Therefore the classical
equation of motion can be obtained by varying action (\ref{eq:35})
which is,

\begin{equation}
z^{\prime}+\frac{2\rho f\left(z\right)}{z}+\rho z^{\prime\prime}-\frac{\rho z^{\prime2}}{2f\left(z\right)}\frac{d}{dz}f\left(z\right)+\frac{2\rho z^{\prime2}}{z}+\frac{z^{\prime3}}{f\left(z\right)}-\frac{1}{2}\rho\left[f\left(z\right)+z^{\prime2}\right]\frac{d}{dz}\phi\left(z\right)=0.\label{eq:36}
\end{equation}
In order to obtain the onshell action, we need to solve the above
differential equation numerically which would however be very difficult.
Fortunately this step can be a little simplified by imposing an additional
constraint as pointed out in \cite{Confining D3 - 2} which in our
setup is,

\begin{equation}
z^{\prime}\left(\rho\right)\big|_{\rho=x}=-\sqrt{f\left(z\right)\left(\frac{e^{\phi}}{\alpha^{2}}-1\right)}\bigg|_{z=z_{0}}.\label{eq:37}
\end{equation}
Altogether, we solve (\ref{eq:36}) numerically with the boundary
condition $z^{\prime}\left(0\right)=0,z\left(0\right)=z_{c}$ and
the constraint (\ref{eq:37}). Then the numerical result for the onshell
action $S$ as a function of $\alpha$ and $q$ is illustrated in
Figure \ref{fig:6} 
 where its exponential behavior $e^{-S}$ is also plotted. The numerical
calculation shows the pair production rate is indeed suppressed by
the presence of the D-instantons which agrees with the potential analysis.
And the the exponential behavior of the classical action approaches
to zero $E_{c}$ obtained by the potential analysis.

\subsection*{The D0-D4 case}

\begin{figure}
\begin{centering}
\includegraphics[scale=0.38]{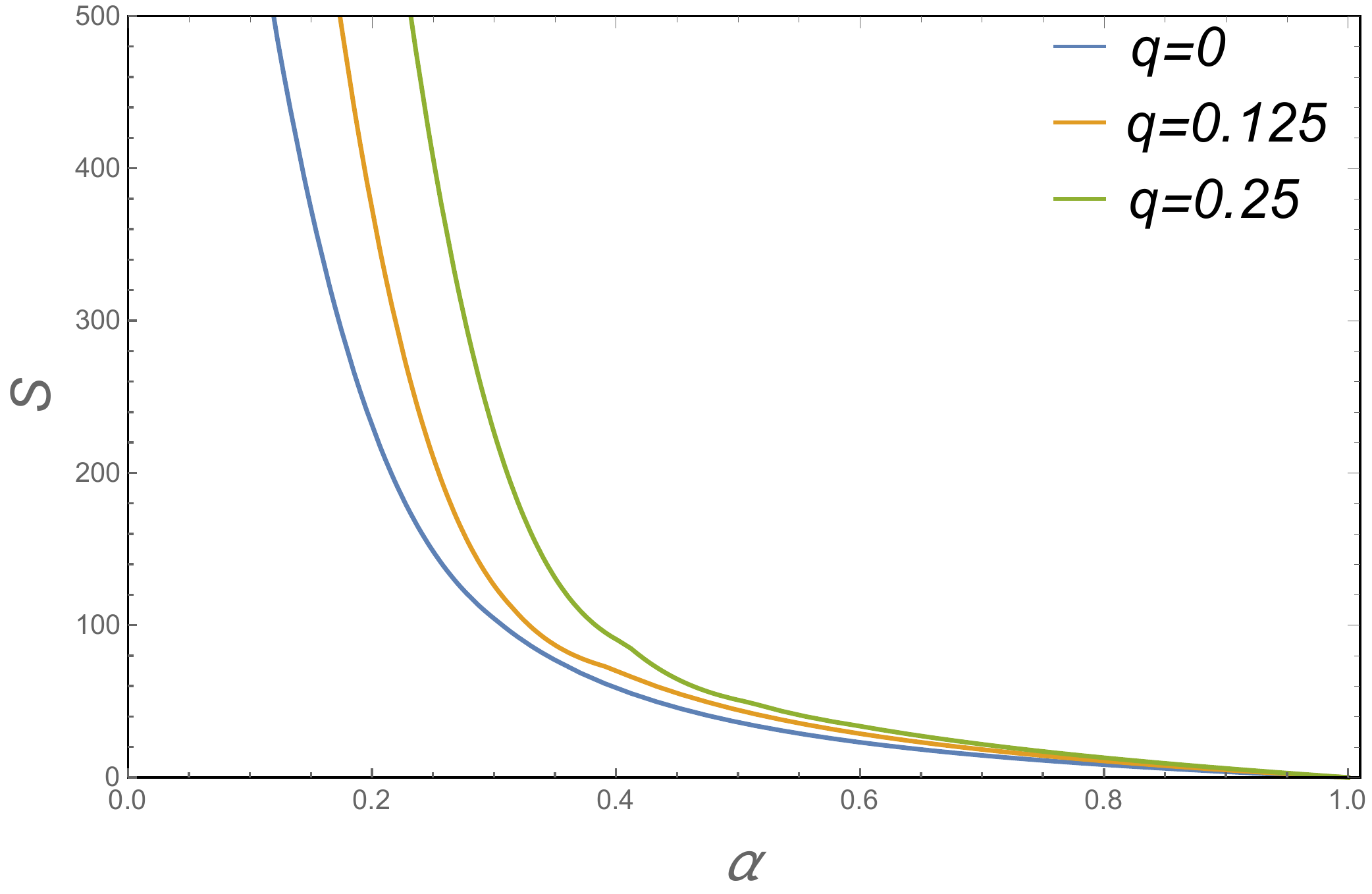}\includegraphics[scale=0.38]{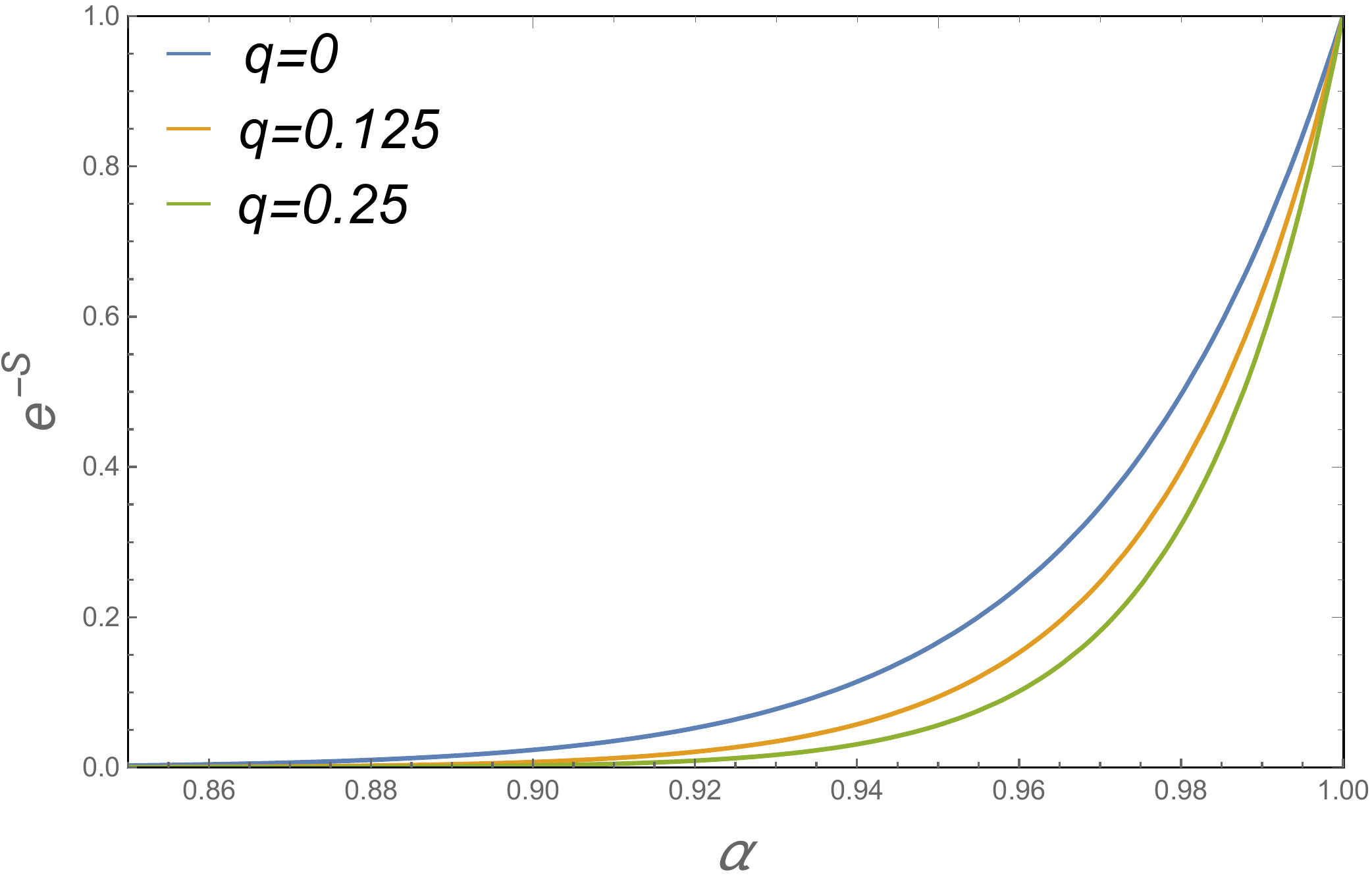}
\par\end{centering}
\caption{\label{fig:7} The classical string action and its exponential behavior
versus $\alpha$ in the D0-D4 background.}

\end{figure}

Let us evaluate the pair production rate in the D0-D4 background.
Following the similar steps as the D(-1)-D3 case, the string action
in the D0-D4 background reads,

\begin{align}
S & =S_{NG}+S_{B_{2}},\nonumber \\
S_{NG} & =16\pi T_{f}R^{3}\int d\rho\frac{H_{0}^{1/2}\left(z\right)\rho}{z^{3}}\sqrt{1+\frac{z^{\prime2}}{f\left(z\right)}},\ f\left(z\right)=1-\frac{z^{6}}{z_{KK}^{6}},\nonumber \\
S_{B_{2}} & =-2\pi T_{f}B_{01}\int_{0}^{x}d\rho\rho=-\pi Ex^{2},\label{eq:38}
\end{align}
where we have imposed the following coordinate transformation,

\begin{equation}
z\left(\rho\right)=\frac{2R^{3/2}}{r^{1/2}}.
\end{equation}
Varying action (\ref{eq:38}), the associated equation of motion is
obtained as,

\begin{equation}
z^{\prime\prime}-\frac{f\left(z\right)+z^{\prime2}}{2H_{0}\left(z\right)}\frac{d}{dz}H_{0}\left(z\right)+\frac{3f\left(z\right)+3z^{\prime2}}{z}-\frac{z^{\prime2}}{2f\left(z\right)}\frac{d}{dz}f\left(z\right)+\frac{z^{\prime3}}{\rho f\left(z\right)}+\frac{z^{\prime}}{\rho}=0.
\end{equation}
Again we solve the above differential equation with the boundary condition
$z^{\prime}\left(0\right)=0,z\left(0\right)=z_{c}$ and an additional
constraint,

\begin{equation}
z^{\prime}\left(\rho\right)\big|_{\rho=x}=-\sqrt{f\left(z\right)\left(\frac{H_{0}\left(z\right)}{\alpha^{2}}-1\right)}\bigg|_{z=z_{0}},
\end{equation}
to obtain the onshell action. Afterwards the classical string action
$S$ and its exponent are numerically evaluated in Figure \ref{fig:7}
. So we see again the pair production rate is suppressed by the presence
of the D-instantons which is in agreement with the potential analysis
in the D0-D4 background.

\section{The electric instability with a single flavor}

The analyse of the Schwinger effect in holography can also be extended
by including flavor \cite{Wenhe,Flavor-1,Flavor-2,Flavor-3}, since
it would also be significant to consider the vacuum instability and
creation rate with the fundamental matters in the Schwinger effect,
which can be introduced when the flavor brane is embedded into the
background geometry. As a supplement to this project, in this section
let us briefly take into account a single flavor brane and investigate
the vacuum instability with an external electric field in the D(-1)-D3
and D0-D4 system respectively.

\subsection*{The D(-1)-D3 case}

To involve the flavors in the dual theory, it is necessary to introduce
the D7-branes as flavors as most approaches in the D3-brane system
\cite{D3D7}. The configuration of various D-branes in the D(-1)-D3
system is given in Table \ref{tab:1}. 
\begin{table}
\begin{centering}
\begin{tabular}{|c|c|c|c|c|c|c|c|c|c|c|c|}
\hline 
 & (-1) & 0 & 1 & 2 & 3 & 4 & 5 & 6 & 7 & 8 & 9\tabularnewline
\hline 
\hline 
D(-1) & - &  &  &  &  &  &  &  &  &  & \tabularnewline
\hline 
D3-brane &  & - & - & - & - &  &  &  &  &  & \tabularnewline
\hline 
D7-brane &  & - & - & - & - & - & - & - & - &  & \tabularnewline
\hline 
\end{tabular}
\par\end{centering}
\caption{\label{tab:1} The configuration of the D-branes in the D(-1)-D3 system.}

\end{table}
 Without loss of generality, we can turn on the components $F_{01}=E,F_{0z},F_{1z}$
of the gauge field strength. Since the electric instability will be
focused, we only need to consider a single D7-brane whose action is
given as,

\begin{equation}
S_{\mathrm{D7}}=-T_{\mathrm{D7}}\int d^{8}xe^{-\phi}\sqrt{-\det\left(g_{\mathrm{D7}}+2\pi\alpha^{\prime}F\right)}-\mu_{7}\int C_{8}.\label{eq:42}
\end{equation}
We note that the supersymmetry is broken down below the energy scale
$M_{KK}$ in our setup. By imposing the background geometry (\ref{eq:2})
into (\ref{eq:42}), the action becomes,

\begin{align}
S & =-2\pi^{2}T_{\mathrm{D7}}V_{4}\int dze^{\phi}\frac{R^{8}}{z^{5}}\sqrt{\xi},\nonumber \\
\xi & =1-\left(2\pi\alpha^{\prime}\right)^{2}\frac{z^{4}}{R^{4}}e^{-\phi}\left[F_{01}^{2}+\left(F_{0z}^{2}-F_{1z}^{2}\right)f\left(z\right)\right],\label{eq:43}
\end{align}
which leads to the following equations of motion for the $U\left(1\right)$
gauge field strength,

\begin{align}
\partial_{z}\left(\frac{f}{z\sqrt{\xi}}F_{0z}\right)=0, & \partial_{0}\left(\frac{f}{z\sqrt{\xi}}F_{0z}\right)=0,\nonumber \\
\partial_{0}\left(\frac{1}{z\sqrt{\xi}}F_{01}\right) & +\partial_{z}\left[\frac{f}{z\sqrt{\xi}}F_{1z}\right]=0.
\end{align}
In particular, when the electric field is static i.e. time-independent,
we can put $\partial_{0}=0$, the above equations of motion reduces
to two constants $j,d$ as,

\begin{equation}
j\equiv2\pi\alpha^{\prime}\frac{f}{z\sqrt{\xi}}F_{1z},d=2\pi\alpha^{\prime}\frac{f}{z\sqrt{\xi}}F_{0z},\label{eq:45}
\end{equation}
which can be interpreted as the electric current and charge in holography
\cite{Wenhe,Flavor-1,Flavor-3}. Plugging (\ref{eq:45}) into (\ref{eq:43}),
we can obtain the effective action as,

\begin{align}
S & =-2\pi^{2}T_{\mathrm{D7}}V_{4}\int dze^{\phi}\frac{R^{8}}{z^{5}}\sqrt{\xi},\nonumber \\
\xi & =\frac{1-\left(2\pi\alpha^{\prime}E\right)^{2}\frac{z^{4}}{R^{4}}e^{-\phi}}{1+e^{-\phi}\frac{z^{6}}{R^{4}f}\left(d^{2}-j^{2}\right)}.\label{eq:46}
\end{align}
Notice that $\xi$ must be positive since the action of stable D-brane
should not admit an imaginary part. Therefore there must be a certain
position $z_{p}$ which changes the sign of the denominator and the
numerator concurrently in $\xi$, otherwise the value of $\xi$ would
becomes negative i.e. the D-brane would be unstable. In this sense
the stable current $j$ can be obtained by solving the following constraints,

\begin{equation}
1-\left(2\pi\alpha^{\prime}E\right)^{2}\frac{z_{p}^{4}}{R^{4}}e^{-\phi\left(z_{p}\right)}=0,1+e^{-\phi\left(z_{p}\right)}\frac{z_{p}^{6}}{R^{4}f\left(z_{p}\right)}\left(d^{2}-j^{2}\right)=0.
\end{equation}
Keeping this in mind, it means in order to study the vacuum instability,
we can consider the situation that the electric field is suddenly
turned on, the vacuum given by $j=0$ would thus become unstable in
the presence of an electric field. In this sense, the above constraints
would not be satisfied as the effective action (\ref{eq:46}) will
now admit an imaginary part. Since the flavor decay rate $\Gamma$
must be proportional to the imaginary part of the effective action,
we can evaluate the vacuum instability by putting $j=0$ in the effective
action which is,

\begin{equation}
\Gamma=\mathrm{Im}S=-2\pi^{2}T_{\mathrm{D7}}V_{4}\int_{z_{KK}}^{z_{*}}dze^{\phi}\frac{R^{8}}{z^{5}}\sqrt{\left(1+e^{-\phi}\frac{z^{6}}{R^{4}f}d^{2}\right)^{-1}\left[\frac{\left(2\pi\alpha^{\prime}\right)^{2}z^{4}}{R^{4}}E^{2}e^{-\phi}-1\right]},
\end{equation}
where $z_{*}$ refers to the position that 

\begin{equation}
\frac{\left(2\pi\alpha^{\prime}\right)^{2}z^{4}}{R^{4}}E^{2}e^{-\phi}-1=\begin{cases}
>0, & z\in\left[z_{*},z_{KK}\right],\\
<0, & z\in\left(0,z_{*}\right].
\end{cases}
\end{equation}
Then the imaginary part of the effective action can be numerically
evaluated and the result is illustrated in Figure. 
\begin{figure}
\begin{centering}
\includegraphics[scale=0.38]{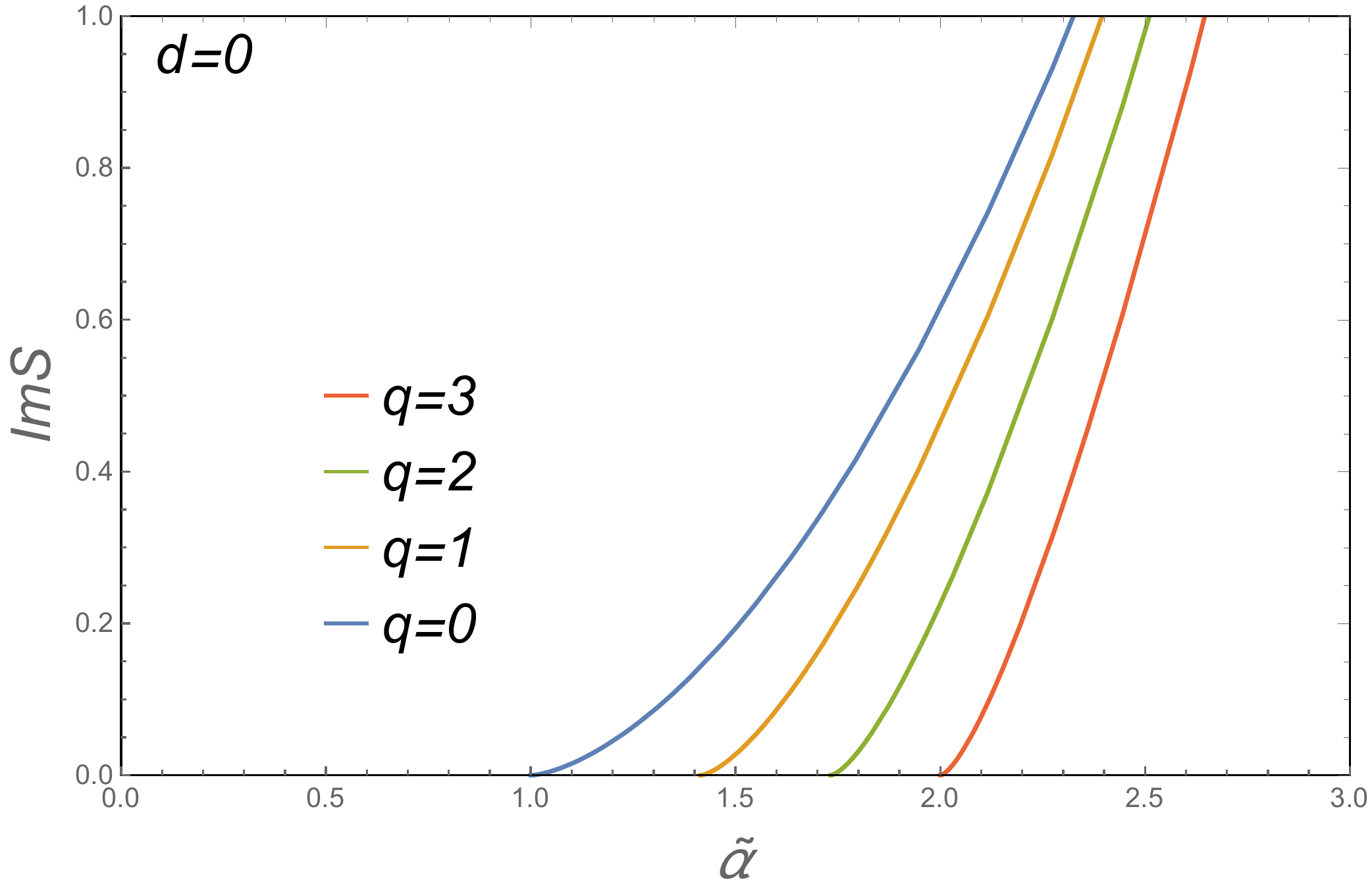}\includegraphics[scale=0.38]{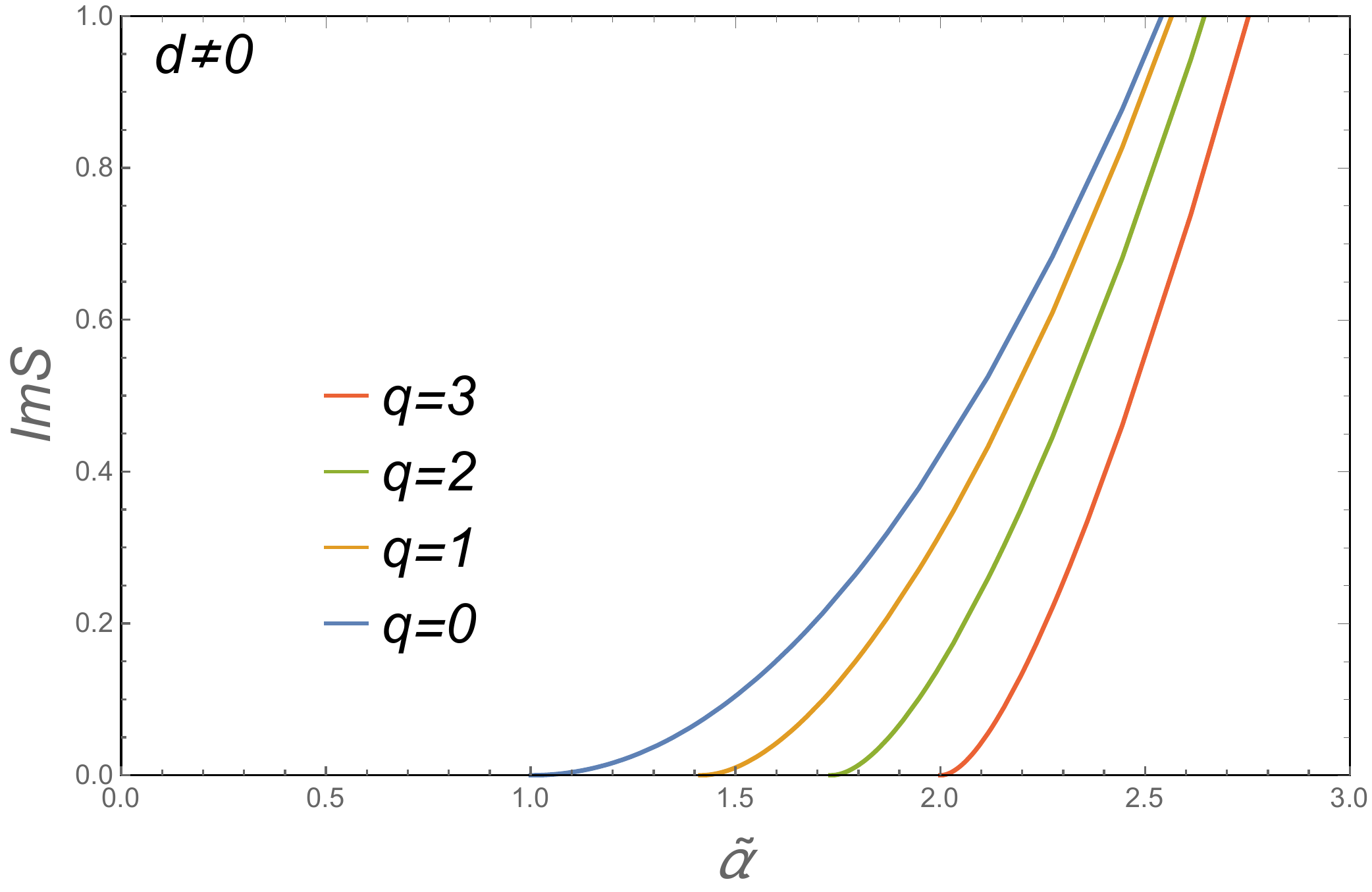}
\par\end{centering}
\caption{\label{fig:8} The imaginary part of the effective action as a function
of $\tilde{\alpha}=E/E_{c}\left(0\right)$ in the D(-1)-D3 background.
We have set $b=1$.}

\end{figure}
 We can see that the decay rate becomes nonzero above the critical
electric field $E_{c}$ and nearly independent on $q$ at sufficiently
large electric field. We also note that the decay rate is suppressed
in the presence of D-instantons which is in agreement with the analysis
in the previous sections.

\subsection*{The D0-D4 case}

\begin{figure}
\begin{centering}
\includegraphics[scale=0.38]{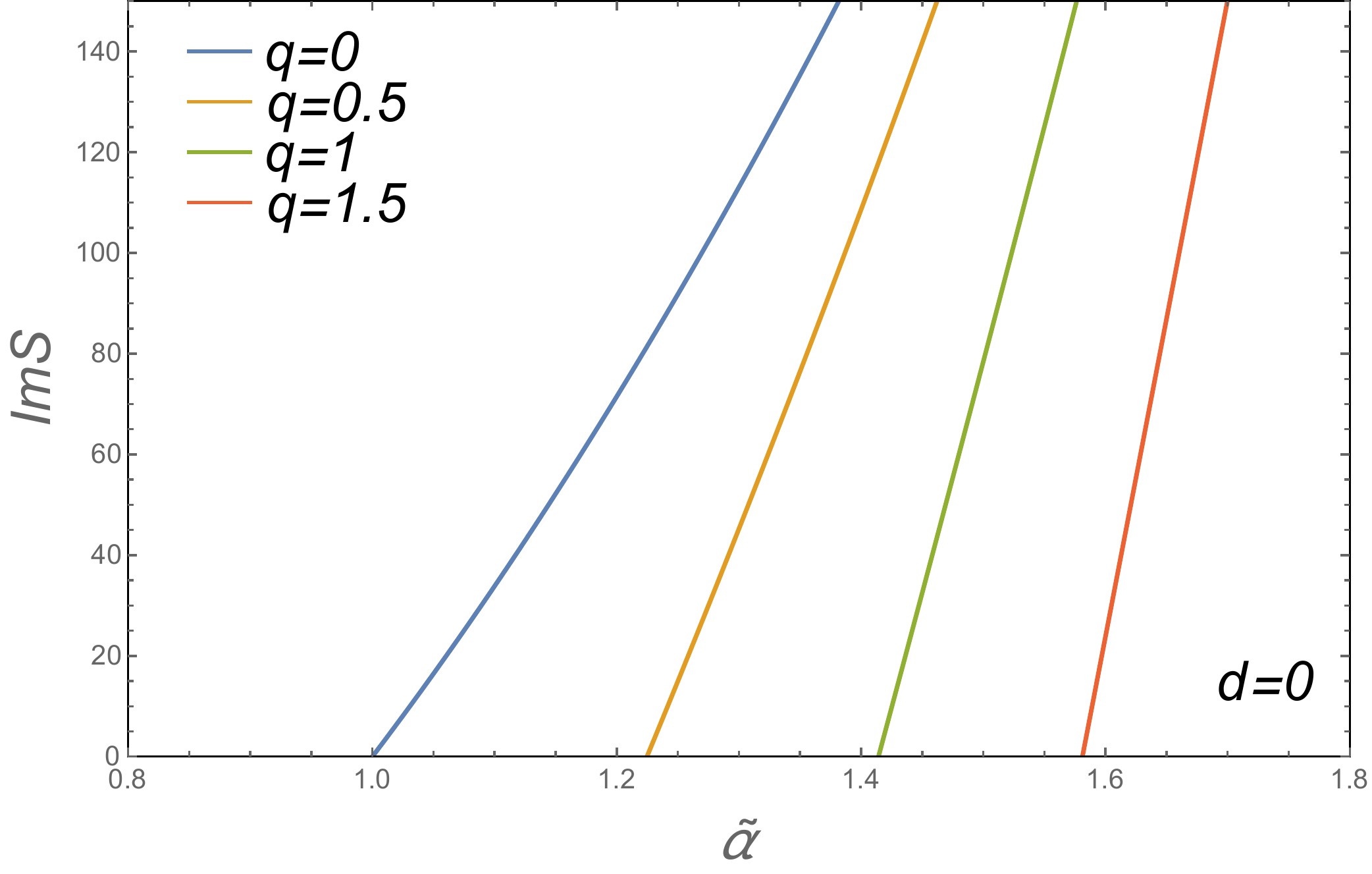}\includegraphics[scale=0.38]{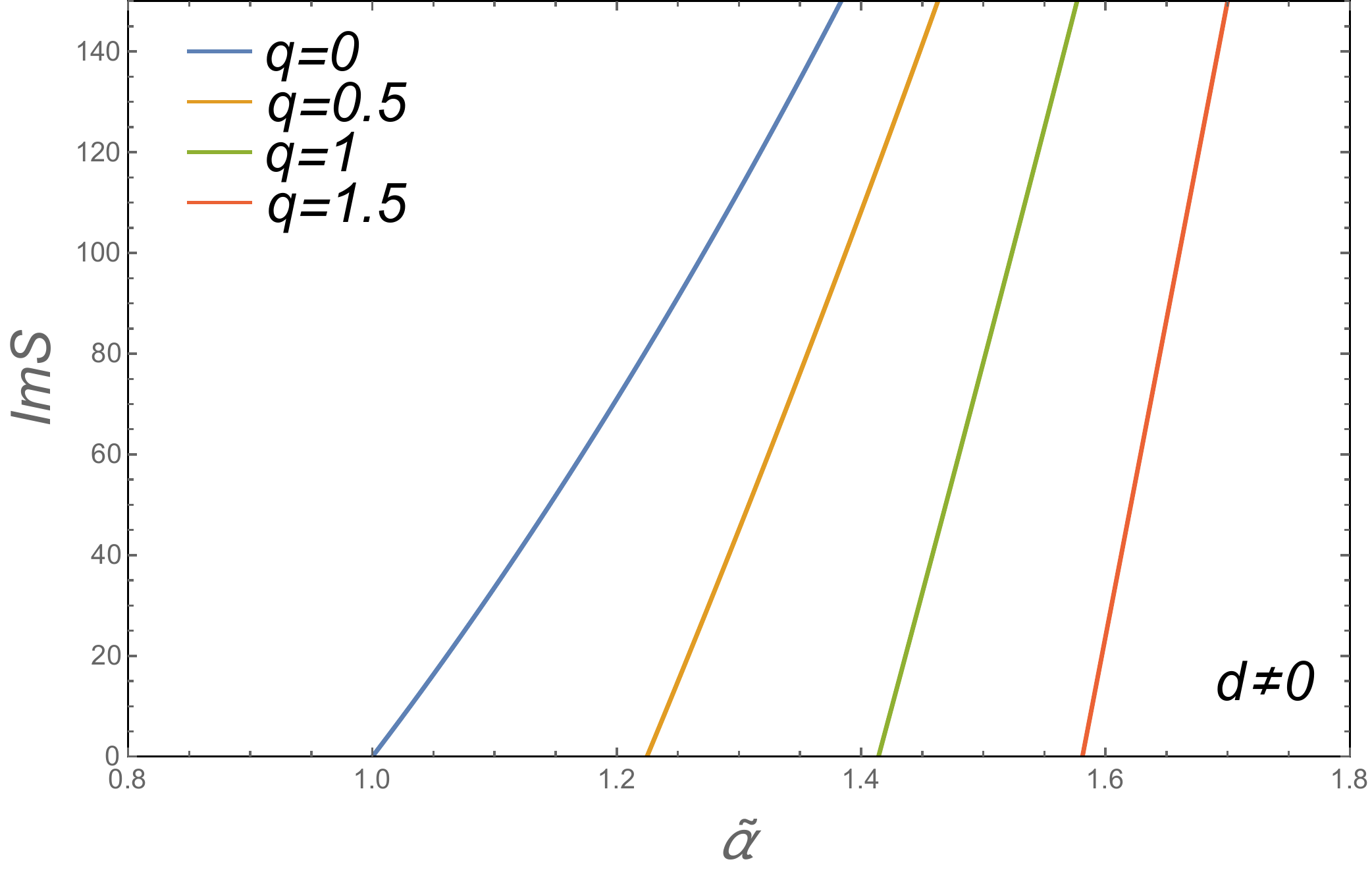}
\par\end{centering}
\caption{\label{fig:9} The imaginary part of the effective action as a function
of $\tilde{\alpha}=E/E_{c}\left(0\right)$ in the D0-D4 background.
The behavior of $\mathrm{Im}S$ is almost not affected by the value
of $d$. We have set $b=1$.}
\end{figure}

In the D0-D4 background, the vacuum instability with an external electric
field can be analyzed by following the discussion in the D(-1)-D3
background while the configuration of the D-branes is distinct as
illustrated in Table \ref{tab:2}. 
\begin{table}
\begin{centering}
\begin{tabular}{|c|c|c|c|c|c|c|c|c|c|c|}
\hline 
 & 0 & 1 & 2 & 3 & 4 & 5$\left(z\right)$ & 6 & 7 & 8 & 9\tabularnewline
\hline 
\hline 
D0-brane &  &  &  &  & - &  &  &  &  & \tabularnewline
\hline 
D4-brane & - & - & - & - & - &  &  &  &  & \tabularnewline
\hline 
D8-brane & - & - & - & - &  & - & - & - & - & -\tabularnewline
\hline 
\end{tabular}
\par\end{centering}
\caption{\label{tab:2} The configuration of the D-branes in the D0-D4 system.}

\end{table}
 The flavor brane is identified as a stack of D8- and anti D8-brane
pair which is vertical to the D4-brane in the $x^{4}-z$ plane. In
this configuration of the D-branes, the DBI action of a single pair
of the flavor brane with nonzero components $F_{01}=E,F_{0z},F_{1z}$
of the gauge field strength can be written as,

\begin{align}
S_{\mathrm{D8}} & =-T_{\mathrm{D8}}\int d^{9}xe^{-\phi}\sqrt{-\det\left(g_{\mathrm{D8}}+2\pi\alpha^{\prime}F\right)}\nonumber \\
 & =-\frac{2^{11}\pi^{2}}{3}T_{\mathrm{D}8}R^{12}V_{4}\int dz\frac{H_{0}^{3/2}}{z^{8}\sqrt{f}}\sqrt{\xi},\label{eq:50}
\end{align}
where

\begin{equation}
\xi=1-\frac{\left(2\pi\alpha^{\prime}\right)^{2}z^{6}}{64H_{0}R^{6}}\left[F_{01}^{2}+f\left(F_{0z}^{2}-F_{1z}^{2}\right)\right].
\end{equation}
Then putting $\partial_{0}=0$ for the static electric field, we obtain
the equations of motion as,

\begin{align}
\partial_{z}\left(\frac{\sqrt{fH_{0}}}{z^{2}\sqrt{\xi}}F_{0z}\right) & =0,\ \partial_{z}\left(\frac{\sqrt{fH_{0}}}{z^{2}\sqrt{\xi}}F_{1z}\right)=0,
\end{align}
which leads to two constants $j,d$ as current and charge, defined
as,

\begin{equation}
j=2\pi\alpha^{\prime}\frac{\sqrt{fH_{0}}}{z^{2}\sqrt{\xi}}F_{1z},d=2\pi\alpha^{\prime}\frac{\sqrt{fH_{0}}}{z^{2}\sqrt{\xi}}F_{0z}.\label{eq:53}
\end{equation}
Imposing (\ref{eq:53}) into (\ref{eq:50}), the effective action
becomes,

\begin{equation}
S_{\mathrm{D8}}=-\frac{2^{11}\pi^{2}}{3}T_{\mathrm{D}8}R^{12}V_{4}\int dz\frac{H_{0}^{3/2}}{z^{8}\sqrt{f}}\sqrt{\left[1+\left(d^{2}-j^{2}\right)\frac{z^{10}}{64H_{0}^{2}R^{6}}\right]^{-1}\left[1-\frac{\left(2\pi\alpha^{\prime}E\right)^{2}z^{6}}{64H_{0}R^{6}}\right]},
\end{equation}
and the stable current $j$ must be obtained by solving the constraints,

\begin{equation}
1+\left(d^{2}-j^{2}\right)\frac{z_{p}^{10}}{64H_{0}^{2}\left(z_{p}\right)R^{6}}=0,\ 1-\frac{\left(2\pi\alpha^{\prime}E\right)^{2}z_{p}^{6}}{64H_{0}\left(z_{p}\right)R^{6}}=0.
\end{equation}
Afterwards we set $j=0$ to investigate the vacuum instability, the
imaginary part of the effective action is therefore obtained as,

\begin{equation}
\mathrm{Im}S=-\frac{2^{11}\pi^{2}}{3}T_{\mathrm{D}8}R^{12}V_{4}\int_{z_{KK}}^{z_{*}}dz\frac{H_{0}^{3/2}}{z^{8}\sqrt{f}}\sqrt{\left(1+d^{2}\frac{z^{10}}{64H_{0}^{2}R^{6}}\right)^{-1}\left[\frac{\left(2\pi\alpha^{\prime}E\right)^{2}z^{6}}{64H_{0}R^{6}}-1\right]},\label{eq:56}
\end{equation}
where $z_{*}$ is given as,

\begin{equation}
\frac{\left(2\pi\alpha^{\prime}E\right)^{2}z^{6}}{64H_{0}R^{6}}-1=\begin{cases}
>0, & z\in\left[z_{*},z_{KK}\right],\\
<0, & z\in\left(0,z_{*}\right].
\end{cases}
\end{equation}
We numerically evaluate (\ref{eq:56}) as a function of $\tilde{\alpha}$
as illustrated in Figure \ref{fig:9}. 
 While the exact behavior of the decay rate is a little different
from the D(-1)-D3 case, our numerical calculation shows again that
the decay rate is nonzero above the critical electric field $E_{c}$
and suppressed by the D-instantons. Obviously the behavior that the
presence of D-instantons suppresses the decay rate is seemingly universal
in the D-brane background with D-instantons.

\section{Summary and discussion}

In this work we have studied the Schwinger effect both in the confining
D3- and D4-brane system with D-instantons. By using the DBI action
of a probe brane, we find the critical electric field $E_{c}$ depends
on the instanton density. Then we perform the potential analysis for
the holographic Schwinger effect, calculate the total potential and
the pair production rate for a pair of particle-antiparticle by taking
into account an electric field. Afterwards we further investigate
the electric instability with a single flavor and evaluate the associate
flavor decay rate. According to our numerical calculation, we find
the critical electric field is in agreement with the analysis of the
DBI action. The presence of instantons increases the potential barrier
and therefore suppresses the pair creation both in the D(-1)-D3 and
D0-D4 approach. Our results agrees well with the evaluation obtained
in the black D(-1)-D3 background at zero temperature limit \cite{deconfined D(-1)-D3}
and the analysis for the flavor brane action in the D0-D4 background
\cite{Wenhe}.

Finally let us give some physical interpretation of our results to
close this work. The D(-1)-D3 case: Since the confining D(-1)-D3 system
holographically corresponds to 3d Yang-Mills plus Chern-Simons theory
below $M_{KK}$, in the dual field the particle may acquire effective
topological mass through the Chern-Simons interaction due to the presence
of the D(-1) branes (D-instantons) which means the particle mass is
increased by the presence of the D-instantons. This would be more
clear if we compute the propagator in the dual theory with a Chern-Simons
term by using the method of quantum field theory, (\ref{eq:4}) which
is,

\begin{equation}
\Delta_{\mu\nu}=\frac{p^{2}\eta_{\mu\nu}-p_{\mu}p_{\nu}-i\kappa g_{YM}^{2}\epsilon_{\mu\nu\rho}p^{\rho}}{p^{2}\left(p^{2}-\kappa^{2}g_{YM}^{4}\right)}+\mathrm{gauge\ term},
\end{equation}
leading to a topological mass $\Delta m=\kappa g_{YM}^{2}$. Therefore
the particle creation in Schwinger effect would be suppressed by the
D-instantons since it is proportional to the factor $e^{-\frac{m^{2}}{E}}$.
Or namely, the barrier of total potential for a pair of particle-antiparticle
is increased by the presence of the D-instantons which is consistent
with what we have obtained in this work. The Schwinger effect with
D-instantons might be very helpful to investigate some phenomena in
3d Maxwell-Chern-Simons theory with electric field e.g. the phase
transition from insulator to conductor in some material.

The D0-D4 case: According to the previous study in this system \cite{WuChao D0-D4,Francesco D0-D4 - 1,Francesco D0-D4 - 2,Li D0-D4,Si-wen Li B1,Si-wen Li B2,Wehe baryon},
the dual field theory is confining Yang-Mills theory with a theta
term (D-instanton density) and the particle mass spectrum is increased
by the presence of the D-instantons (D0-branes). Hence the particle
creation in Schwinger effect should also be suppressed or namely the
total potential barrier is increased by the instantons which is consistent
with the holographic analysis in this work. The total potential with
D-instantons in Schwinger effect would be also remarkable to study
the P or CP violation in QCD, especially in the heavy-ion collision,
since an extremely strong electromagnetic field would be generated.
On the other hand, in the collision, there might be a metastable state
with nonzero vacuum theta angle produced in the hot and dense condition
when the deconfinement phase transition happens \cite{Kharzeev,Buckley}.
Therefore the metastable state may be excited by the strong electric
field through the Schwinger effect and the particle pair creation
in the heavy-ion collision would be affected by the theta angle (D-instantons),
reflecting the behaviors of the total potential. So Schwinger effect
with D-instantons might be an observable phenomenon to confirm whether
P or CP violation in QCD exists.

\section*{Acknowledgements}

This work is supported by the research startup foundation of Dalian
Maritime University in 2019 under Grant No. 02502608, the Fundamental
Research Funds for the Central Universities under Grant No. 3132021205,
the National Natural Science Foundation of China (NSFC) under Grant
No. 12005033 and Grant No. 11947008.

\end{document}